\documentclass[a4paper,BCOR1cm,12pt,headsepline, normalheadings, bibtotoc]{scrbook}
\usepackage[ansinew]{inputenc}
\usepackage{graphicx, epsfig,float,bbm}

\newcommand{\be}{\begin{equation}}
\newcommand{\ee}{\end{equation}}
\newcommand{\ba}{\begin{eqnarray}}
\newcommand{\ea}{\end{eqnarray}}

\newcommand{\beq}{\begin{equation}}
\newcommand{\eeq}{\end{equation}}
\newcommand{\beqa}{\begin{eqnarray}}
\newcommand{\eeqa}{\end{eqnarray}}

\newcommand{\openone}{\ensuremath{\mathbbm{1}}}

\begin{document}

\title{The Foundations of\\Quantum Information and\\ Feasible Experiments}
\author{Christoph Simon\\ \\
\large Dissertation zur Erlangung des akademischen Grades\\
\large Doktor der Naturwissenschaften\\
\large an der Fakult\"{a}t f\"{u}r Naturwissenschaften und Mathematik\\
\large der Universit\"{a}t Wien}
\date{\large Wien, im Dezember 2000 \\[30ex]
\normalsize Gef\"{o}rdert vom Fonds zur F\"{o}rderung der wissenschaftlichen Forschung,\\
\normalsize Projekte Nr. S6503 und F1506}
\maketitle

\frontmatter

\tableofcontents
\chapter{Abstract}

This thesis contains results on different questions in quantum
information. It consists of four chapters. The subject of the
first chapter is the copying of quantum states by stimulated
emission. According to the no-cloning theorem by Wootters and
Zurek\cite{wzurek} it is fundamentally impossible to build a
machine which would be able to produce an exact copy of a quantum
system in an unknown state. The impossibility of perfect copying
follows immediately from the linearity of quantum physics.
Approximate copying however is compatible with the principles of
quantum mechanics. Quantum mechanics only gives bounds on the
fidelity of the copies. Stimulated emission, which is at the heart
of the laser, is a natural candidate for the practical realization
of a quantum copier. Here it is shown that optimal (i.e.
saturating the quantum mechanical bounds) copying of photons can
be realized by stimulated emission in simple quantum optical
systems, for example three-level atoms. The fidelity of the copies
is limited by the unavoidable presence of spontaneous emission,
which thus assures that the quantum mechanical bounds are obeyed.

In spite of its non-local features such as the violation of Bell's
inequalities, quantum physics is entirely compatible with the special
theory of relativity. In particular, entangled states cannot be used for
superluminal communication. This peaceful co-existence has led to the
question whether the impossibility of superluminal signaling could be used
as an axiom in deriving basic features of quantum mechanics from
fundamental principles. In the second chapter we show that this is indeed
the case. If the kinematical features of quantum
physics including the projection postulate
are assumed to be
given, then its dynamical rules can be derived with the help of the
no-signaling condition. This also puts constraints on possible non-linear
modifications of quantum mechanics.

Quantum mechanics usually only allows statistical predictions for
the behavior of individual physical systems. The third and fourth
chapter of this thesis are devoted to theorems on the existence of
hidden variables which would make it possible to make predictions
for individual systems. One of the classical hidden-variable
theorems is the one by Kochen and Specker, which states that so
called non-contextual hidden variables are incompatible with
quantum mechanics. A new, much simplified, version of this theorem
is given, which leads to a proposal for a simple experimental test
of non-contextual hidden variables, for example with single
photons and linear optical elements.

The fourth chapter treats the derivation of hidden-variable
theorems for real experiments, in particular for finite
measurement precision. This investigation was motivated by recent
claims that the Kochen-Specker theorem loses its validity under
such conditions. It is shown that the basic statements of hidden-variable
theorems are robust under real-world conditions.

\mainmatter

\addchap{Preface}

Let me begin with some remarks on how I ended up doing precisely
the things which are collected in this thesis, and not something
else. The emphasis of my undergraduate studies in Vienna and then
also in Paris was on theoretical particle physics. I had chosen
this subject at the beginning of my studies because it seemed the
most fundamental area of physics.

But already as an undergraduate I became very fascinated by the
mysterious features of quantum mechanics. I remember that I
first heard about Bell's inequalities from Robin Michaels, then a
mathematics student in Cambridge, shortly after beginning my
studies. I already knew the basic principles of quantum mechanics
at that time, but I had had a naive realistic view concerning its
statistical predictions, as I realized through our discussion.
Later, I attended the seminar on the Foundations of Quantum
Mechanics organized by Reinhold Bertlmann and Anton Zeilinger,
which was my first contact with Anton.

Towards the end of my studies I realized that I wanted to learn
more about the fundamental questions of quantum mechanics, and, if
possible, work on them. I thought that the most fundamental
question was whether there is something beyond quantum mechanics,
or whether we have to content ourselves with its highly
idiosyncratic ways of giving us information about the world. I was
aware that I would most probably not be able to answer this
question during my PhD, but it certainly was the guiding star of
my decision. I was also aware that the question is an experimental
one. Up to this point, I had not learned very much about
experiments.

Given all this, it was quite natural that I joined Anton's group
in Innsbruck in December 1997. It is worth mentioning that I
arrived there more or less simultaneously with the first TV crews
wanting to know about teleportation. I was determined to learn as
much as possible about the experimental side. I was lucky enough
to spend the first year working as an apprentice on the up to now
best experimental test of Bell's inequalities, Gregor Weihs' PhD
experiment. Although my time as an experimentalist was actually
not very long, it was a very valuable experience. I am convinced
that it made me a better physicist, also in theory.

I had always intended to do some theoretical work on the side. Our
move from Innsbruck to Vienna created some additional spare time.
Both my work on cloning and on a Kochen-Specker experiment was
started in 1998, while we were still in Innsbruck. Anton proposed
a Kochen-Specker experiment as a possible topic for my thesis
shortly after my arrival in Innsbruck, and he also suggested the
work of Cabello and Garc\'{\i}a-Alcaine as a starting point. I still
remember the first discussions with Marek \.{Z}ukowski,
Harald Weinfurter and Anton quite vividly. We finished this project much
later, in March 2000.

My interest in cloning was triggered at a European Quantum
Information meeting in Helsinki. Gregor told me that in Innsbruck
they had been discussing the relation of cloning and stimulated
emission before, which started our common work on this subject. I
remember that from the beginning Anton was interested in the
relation between cloning and superluminal communication. The year
after that we continued (and in a sense completed) our work
together with Julia Kempe.

During my undergraduate studies I had not heard much about the new
field of quantum information. I remember reading an introduction
to quantum computing (by Adriano Barenco) in Paris,
which I had downloaded from the
quant-ph folder. In Innsbruck, somewhat unexpectedly, I found
myself in one of the centers of the new field. There was not only
our group but also our theoretical and experimental colleagues,
with Peter Zoller, Ignacio Cirac and Rainer Blatt, so there was a lot
to learn.

Today, when asked what I do, I often call myself a theorist in
quantum information, and I will soon be a postdoc in a ``centre
for quantum computation''. I have remained true to my foundational
interests, as illustrated by my work with \v{C}aslav Brukner and
Anton on hidden-variable theorems for real experiments, and my
work with Vladimir Bu\v{z}ek and Nicolas Gisin on the no-signaling
condition. It was of course very helpful to have a boss like
Anton, who is himself so deeply fascinated by quantum physics. I
am curious to see which unexpected turns the future will bring.
Meanwhile I hope that the products of my efforts collected here
will be of interest to some of my colleagues.

\chapter{Cloning via Stimulated Emission}

\section{Introduction -- Quantum Information}

The first part of this thesis is concerned with quantum cloning,
in particular with the realization of quantum cloning using
stimulated emission. As our work on this topic was usually
published under the heading of quantum information in the
respective journals it seems appropriate to start with a few
remarks on quantum information in general \cite{qi}.

What is quantum information? Let me try to give two tentative
definitions, both formulated as questions. The first one may seem
rather too broad, the second one rather too narrow. The first runs
as follows: what can you do with quantum mechanics that you cannot
do classically? This describes the spirit of the field rather
well, but it doesn't quite explain the name quantum information.
The second runs: what happens if one has qubits instead of
classical bits? As a definition of quantum information this is
certainly too restricted, but some of its proudest achievements,
such as the celebrated quantum computing algorithms, fit very well
into this framework. Furthermore, this definition clearly
emphasizes the information processing aspect of the field.

Nowadays, everybody knows what a bit is. Physically a bit is
represented by a system with two possible states which are clearly
distinguishable, conventionally denoted as 0 and 1. There are of
course many possible physical implementations, from smoke signals
over pulses of voltage to zones of magnetization on a hard disk. A
common feature of all these implementations is that the system can
always be determined to be in one of the two relevant states, 0 or
1. If this is not possible, we are dealing with a bad
implementation. Now consider a quantum system, which can also be
in two clearly distinguishable, i.e. orthogonal states, $|0\rangle$  and $|1\rangle$.
Then it follows from the basic principles of quantum mechanics
that all properly normalized superposition states

\be
\alpha\vert0\rangle+\beta\vert 1 \rangle \mbox{, with }\vert\alpha\vert^2+\vert\beta\vert^2=1
\ee

are also possible physical states of the system. In the following
we will denote such two-dimensional systems as qubits.

The distinction between qubits and classical bits becomes even more pronounced
when several systems are considered. Two classical bits can be in 4 different
states: 00, 01, 10 and 11. For qubits any linear combination of the
corresponding basis states, $|00\rangle$, $|01\rangle$, $|10\rangle$, and
$|11\rangle$, corresponds to a possible state of the physical system.  This
means that the two qubits can also be in entangled states, such as

\be
|\psi\rangle=\frac{1}{\sqrt{2}}(|00\rangle+|11\rangle). \ee

Entangled states have been known for a long time to exhibit
phenomena which are entirely incompatible with the world-view of
classical physics, such as the violation of Bell's inequalities
\cite{azent,bell}.
It may therefore not seem too surprising that replacing bits by
qubits can lead to rather different new results.

The most important theoretical developments in the field of
quantum information have been the discovery of quantum
cryptography and of the quantum computing algorithms by Shor \cite{shor} and
Grover \cite{grover}.
Quantum cryptography \cite{crypto} establishes an entirely secure
communication channel between two distant parties. It is based on
the fact that in quantum mechanics in general there is no way of
performing a measurement without disturbing the state of the
system. This implies that in an appropriately designed scheme, any
eavesdropper trying to listen in will always be detected. There is
no parallel to quantum cryptography in the realm of classical
physics. In the past few years there have been many quantum
cryptography experiments of increasing practicality and
sophistication \cite{jennewein,naik,tittel}. Of all the quantum information paradigms, quantum
cryptography has certainly come closest to being a usable
technology.

The first milestone in the field of quantum computation, i.e.
computation based on qubits instead of classical bits, was Shor's
discovery of his factoring algorithm \cite{shor}. Shor showed that on a
quantum computer it is possible to factor large numbers by
performing a number of operations that scales only polynomially
with the size of the input, whereas the fastest known classical
algorithms require an exponential number of steps. This was the
first serious indication that in the long run quantum computers
may be dramatically faster than classical computers. Since then a
lot of work has been invested in trying to find other relevant algorithms
which show a similar exponential speedup. So far these
attempts have not been successful. However, Grover \cite{grover} succeeded
in showing that quantum computers also out-perform classical
computers in solving a very common task, namely in searching a
completely unordered database - imagine for example that you are
given a phone number and telephone directory and you want to find
out which name the number belongs to. Although the speedup is
less dramatic in this case than in the case of factoring, Grover's
discovery may well turn out to be of great practical importance.
It should be noted, however, that it requires the database to be given
in the form of a quantum mechanical superposition state.

As the existence of entangled states is one of the most important
differences between quantum and classical mechanics, it is not too
surprising that in both these quantum algorithms entanglement
plays a decisive role.

As for the practical implementation of quantum computing, there is still a long
way to go. However, simple algorithms using a small number of qubits have been
realized in various physical systems, such as in nuclear magnetic resonance
\cite{nmr}, cavity QED \cite{cqed,flythru} and ion traps \cite{cirac,monroe}.
It is fair to say that nobody currently knows whether it
will be possible to build a large-scale quantum computer working with hundreds
or thousands of qubits.  The main practical difficulty is
decoherence \cite{decoherence}.  Under normal circumstances, a multi-qubit
entangled state would immediately be destroyed by the interaction with its
environment. It is virtually impossible to shield a quantum computer from its
environment to such a degree as to prevent this from happening. However, there
is hope, based on the development of quantum error correction
\cite{shor2,cald,steane} and fault
tolerant quantum computation \cite{faultt}. Using methods that are similar to classical
error-correction techniques, but much more subtle, it is possible to detect and
correct errors occurring during the quantum computation, including those caused
by decoherence. However, these schemes already presume the existence of quantum
computer elements which work quite reliably, that is, where the error
probabilities are below certain threshold values, typically far below
the percent level (per operation).

Let us now come back to the differences between classical bits and
qubits. One very important difference is the following: The state
of an unknown bit is very easy to determine because it can only be
0 or 1. On the other hand it is completely impossible to determine
the state of an unknown qubit if one is given only a single copy
of the system. This is easy to see: the unknown qubit could be in
any superposition state
$\alpha |0\rangle + \beta |1\rangle$
and it turns out that the best thing that the observer can do is
to perform a projective measurement in some basis of the Hilbert
space spanned by the states $|0\rangle$ and $|1\rangle$ \cite{massarpop}.
Such a measurement only gives
him one bit of information while he would need an infinity of bits
to exactly determine $\alpha$ and $\beta$. Interestingly, it is possible
to teleport an unknown qubit to a distant location with the help
of quantum entanglement \cite{teleport,telep}.
This can be done without finding out anything about the qubit's state.

A related distinction between classical bits and qubits is the
impossibility of copying the latter. It is very easy to copy a
classical bit: even if it is originally unknown, one simply has to
determine its state, 0 or 1, and then produce one more bit in the
desired state. Our discussion above indicates that such an
approach cannot work for qubits. Actually there is no way of
constructing a perfect quantum copying machine, that is, a machine
which given a qubit in an unknown state produces a copy. This is
the content of the famous quantum no--cloning theorem
\cite{wzurek,dieks}. In the
following we will see that the impossibility of copying quantum
information has deep roots: it is related to the linearity of
quantum mechanics, which is in turn related to the impossibility
of superluminal communication.

\section{Signaling and Cloning}\label{sandc}
To our knowledge, the discussion about the cloning of quantum
systems started with a paper by Herbert \cite{herbert}, where he
proposed a method for superluminal communication. His scheme made
use of pairs of entangled particles shared by the two parties that
would like to communicate (Alice and Bob), and of what he called
idealized laser tubes, which would today be called universal
cloning machines. The basic idea of his proposal was the
following. Alice and Bob each have one member of a pair of
entangled particles, e.g. photons described by the state
\begin{equation}
|\psi\rangle=\frac{1}{\sqrt{2}}(|V_A H_B\rangle - |H_A V_B\rangle),
\end{equation}
where $V$ and $H$ denote vertical and horizontal polarization
respectively. Alice can measure the polarization of her particle
either in the basis $|V\rangle, |H\rangle$ or in the basis
$|P\rangle, |M\rangle$, where
$|P\rangle=\frac{1}{\sqrt{2}}(|V\rangle+|H\rangle)$ and
$|M\rangle=\frac{1}{\sqrt{2}}(|V\rangle-|H\rangle)$. If Alice
measures in the $V/H$ basis and finds $|V\rangle$ ($|H\rangle$),
Bob's photon is reduced to $|H\rangle$ ($|V\rangle$), while if she
measures in the $P/M$ basis (results $|P\rangle$ and $|M\rangle$)
Bob's photon is reduced to the corresponding states in that basis
($|M\rangle$ and $|P\rangle$ respectively). Although the states on
his side are therefore different depending on Alice's choice of
basis, a priori this does not allow Bob to know her choice because
he cannot discriminate $|V\rangle$ from $|P\rangle$ by a single
measurement (having only a single copy). But imagine that he has a
machine that can produce an arbitrary number of copies of any
one-photon state, or at least of the states $|V\rangle$ and
$|P\rangle$. This would allow Bob to discriminate the two states and in
this way Alice's two choices of basis. If Bob's copier works fast
enough, this establishes a superluminal communication channel.
Herbert proposed stimulated emission as a possible working
principle for his copying machine. He was aware of the fact that
spontaneous emission could be a problem, but thought that it would
not be fatal for the scheme.

On the other hand, it is possible to show in a general way that
superluminal signaling (or, more precisely, signaling between systems whose
operator algebras commute) is not possible in
quantum mechanics \cite{ghirardi}. One way of explaining why Alice
cannot signal to Bob is the following. Bob does not know which
result Alice got, therefore he has to trace over her
degrees of freedom. Thus his photon is described by a density matrix
$\frac{1}{2}(|V\rangle\langle V|+|H\rangle\langle H|)$, if she
measured in the $V/H$ basis, and $\frac{1}{2}(|P\rangle\langle
P|+|M\rangle\langle M|)$, if she measured in the $P/M$ basis. Of
course, these two density matrices are identical, so there is no
way for him to tell what she did. Now it is clear that no
quantum--mechanical device can lead to a distinction between
identical density matrices. So something has to be wrong
with Herbert's argumentation. Cf. also our discussion in the second chapter
of this thesis.

Wootters and Zurek \cite{wzurek} and Dieks \cite{dieks} showed that the problematic part
of Herbert's proposal is the copying procedure.
To see this consider a device which produces perfect copies of $V$ and $H$ polarized photons,
i.e. which performs the following unitary transformation:
\ba\label{perfclone} |V\rangle|\psi_0\rangle \rightarrow |V\rangle|V\rangle| \psi_V\rangle
\nonumber\\
| H \rangle|\psi_0 \rangle \rightarrow |H\rangle|H\rangle|\psi_H\rangle \ea

How does such a device act on a photon which is linearly polarized
under $45$ degrees, i.e. described by a state vector
$|P\rangle=\frac{1}{\sqrt{2}}(|V\rangle+|H\rangle)$? From a
perfect copier we would expect an output of the form
$|P\rangle|P\rangle|\psi_P\rangle$. On the other hand it follows
from Eq. (\ref{perfclone}) that \be
\frac{1}{\sqrt{2}}(|V\rangle+|H\rangle)|\psi_0\rangle\rightarrow\frac{1}{\sqrt{2}}\left(|V\rangle|V\rangle|\psi_V\rangle+
|H\rangle|H\rangle|\psi_H\rangle\right)\ee It is not hard to see
that this output will never be of the desired form. If
$|\psi_V\rangle\neq|\psi_H\rangle$, the state of the two copies
will even be mixed, which is certainly not what we want. If
$|\psi_V\rangle=|\psi_H\rangle$, the state of the two copies will
be given by $\frac{1}{\sqrt{2}}\left(|V\rangle|V\rangle+
|H\rangle|H\rangle\right)$ which is different from the desired
state
$|P\rangle|P\rangle=\frac{1}{2}(|V\rangle+|H\rangle)(|V\rangle+|H\rangle)$.
This shows that a perfect cloner of photons in the $V$/$H$ Basis
is a maximally bad cloner for states in the complementary basis.
Perfect cloning of general input states is impossible. This is the
famous no--cloning theorem. It was
also pointed out by Mandel \cite{mandel} and Milonni and Hardies \cite{milonni},
that perfect cloning in stimulated--emission
schemes such as proposed by Herbert is prevented by the
unavoidable presence of spontaneous emission. This will be
discussed in detail in the sequel. With all these results the
discussion on cloning was closed for about fourteen years.

In 1996 Bu\v{z}ek and Hillery \cite{buzekhillery} considered cloning from a
different point of view. They proposed and studied an approximate copying
machine, i.e. a device which, given a qubit in an unknown input state, produces
two approximate copies. They also demanded that their machine be universal,
i.e. the quality of the copies should be the same for all inputs.
Bu\v{z}ek and Hillery showed that such a universal quantum copying machine can be described
by the following unitary transformation:
\ba\label{bhcloner} |0_a\rangle|0_b\rangle|0_c\rangle \rightarrow \sqrt{\frac{2}{3}} |0_a\rangle|0_b\rangle|1_c\rangle+
\sqrt{\frac{1}{6}} (|0_a\rangle|1_b\rangle+|1_a\rangle|0_b\rangle) |0_c\rangle \nonumber\\
|1_a\rangle|0_b\rangle|0_c\rangle \rightarrow \sqrt{\frac{2}{3}} |1_a\rangle|1_b\rangle|0_c\rangle+
\sqrt{\frac{1}{6}} (|0_a\rangle|1_b\rangle+|1_a\rangle|0_b\rangle) |1_c\rangle
\ea
where $a$ is the system to be copied, $b$ is an auxiliary qubit which becomes a
copy of $a$ through the transformation, and $c$ is another auxiliary qubit.
Several remarks are in order.  The final state of Eq. (\ref{bhcloner}) is
invariant under the exchange of $a$ and $b$, i.e.  the copying machine produces
two copies with exactly the same properties. In this respect, the distinction
between the original qubit and its copy is completely lost during the copying
procedure. The quality of the copies can be quantified by the fidelity with
respect to the input state, i.e. by $ F=\langle\psi|\rho_a|\psi\rangle$, where $|\psi\rangle$ is
the state of the original qubit and $\rho_a=\mbox{Tr}_{bc}\rho_{abc}$ is the reduced
density matrix of the qubit $a$ in the final state $\rho_{abc}$. Applying this formula to
Eq. (\ref{bhcloner}) gives $F=\frac{5}{6}$.  One can show that the transformation
(\ref{bhcloner}) is indeed universal, i.e.  any arbitrary input state
$\alpha|0\rangle+\beta|1\rangle$ is copied with the same fidelity. In their seminal paper
Bu\v{z}ek and Hillery did not show that their copying machine is optimal.

The Bu\v{z}ek-Hillery construction was generalized by Gisin and
Massar \cite{gisinmassar}, who found a cloning transformation producing $M$ copies
starting from $N$ identical qubits. It is described by \ba
\label{gm} & & U_{N,M}|N_\psi\rangle=\sum_{j=0}^{M-N}\alpha_j|(M-j)\psi,j\psi^\bot\rangle\otimes
R_j(\psi),\nonumber\\
& & \alpha_j=\sqrt{\frac{N+1}{M+1}}\sqrt{\frac{(M-N)!(M-j)!}{(M-N-j)!M!}}
\ea
where $|N\psi\rangle$ is the input state consisting of $N$ qubits all in the state $\psi$,
we have denoted $|(M-j)\psi,j\psi^\bot\rangle$ the symmetric and normalized state
with $M-j$ qubits in the state $\psi$ and $j$ qubits in the orthogonal state $\psi^\bot$.
$R_j(\psi)$ are orthogonal states of the ancillary qubits which can be written as
$R_j(\psi)=|(M-1-j)\psi^\ast,j(\psi^\ast)^\bot\rangle$,
where $\psi^\ast$ is the complex conjugate
of $\psi$.

The transformation (\ref{gm}) looks rather intimidating, or at least
unintuitive. We will see that actually it arises quite naturally
in the context of stimulated emission for appropriately designed
systems. The form of the coefficients $\alpha_j$ was also already
explained in a rather intuitive way by the work of Werner \cite{werner}, which
we will discuss below. The optimality of the Bu\v{z}ek-Hillery and
Gisin-Massar transformations was first shown by Bru\ss{} and
co--workers in \cite{brussprl}, who made use of the relations between
quantum cloning and the estimation of quantum states. The general
idea of this approach is the following: One way of estimating an
unknown quantum state is to first clone it and then perform a
state estimation of its clones. But this cannot be better than the
optimal state estimation, whose fidelity is known for qubits. On
the other hand, one way of cloning an unknown state is to first
estimate it and then produce as many copies as desired of the
estimated state. But this cannot be better than optimal cloning.
Using these two relations, Bru\ss{} et al. derived bounds on the
optimal cloning fidelities which are saturated by the
Bu\v{z}ek-Hillery and Gisin-Massar transformations. The optimal
fidelity for the case of $N$ to $M$ cloning of qubits is \be
F_{N\rightarrow M}=\frac{NM+M+N}{M(M+2)}\ee

It is not hard to see that the Gisin--Massar transformation (\ref{gm})
has this copying fidelity. In this case the fidelity can be
expressed as the mean relative frequency of qubits in the original
state $\psi$ in the final state, i.e. \be
F=\sum^{M-1}_{j=0}\frac{M-j}{M}\alpha^2_j=F_{N\rightarrow M},
\ee
as can be confirmed using the explicit form of the coefficients
$\alpha_j$. The Bu\v{z}ek--Hillery transformation is contained in
the Gisin--Massar transformation as the simplest special case. As
mentioned before it leads exactly to $F=F_{1\rightarrow 2}=\frac{5}{6}$.

Bu\v{z}ek, Hillery and Werner showed that by choosing the states
$R_j(\psi)$ in the Gisin--Massar transformation (\ref{gm}) appropriately, it
is possible to realize optimal universal cloning and the optimal
universal NOT operation simultaneously \cite{buzeknot}. The ideal universal NOT
would be an operation that produces the orthogonal complement of
an arbitrary qubit. Like perfect cloning, this is prohibited by
quantum mechanics.

The understanding of the structure of the cloning transformations
was significantly deepened by the work of Werner, who proved that
the final density matrix of the clones in the optimal $N$ to $M$
cloning transformation can be found in the following way, apart
from normalization:
\be\label{werner}
\rho_M=P^+_M(\sigma^{\otimes N}\otimes \openone^{\otimes (M-N)})P^+_M
\ee
where $\sigma=|\psi\rangle\langle\psi\vert$, is the original state of the $N$
initial qubits, $\openone$ is the completely mixed density matrix, and
$P^+_M$
is the projector onto the completely symmetric subspace of the $N$--qubit
Hilbert space. This is a rather intuitive formula. One can
say that at the beginning all the information is contained in the
$N$ original qubits. The $M-N$ auxiliary qubits are completely
mixed and thus contain no information. Then, the information is
distributed over all $M$ qubits in a completely symmetric way. The
expression (\ref{gm}) for the coefficients $\alpha_j$ can be easily derived
from Eq. (\ref{werner}).
\section{Optimal Universal Cloning Transformations}\label{trafo}

The consideration of Werner's formula Eq. (\ref{werner}) leads to a better
understanding of the Bu\v{z}ek--Hillery and Gisin--Massar transformations.
Let us start with the case of $1\rightarrow 2$ cloning of qubits.
We look for a universal transformation producing two approximate copies
of the state $\psi$. It is clear that ancillas may be needed. What are the
basic elements out of which the final state could be built? If, for the moment,
we take universality to mean that no direction may be singled out by the form of
the final state, apart from the direction of $\psi$, then it is clear that
the only elements permitted are $\psi$ itself, which may appear only linearly
(because of the linearity of quantum mechanics), and the singlet state
of two qubits $S=\frac{1}{\sqrt{2}}\left(\vert0\rangle \vert1\rangle-\vert1\rangle\vert0\rangle    \right)$. The singlet can be rewritten as $S=\frac{1}{\sqrt{2}}\left(\psi \psi^{\perp} - \psi^{\perp}\psi\right)$,
where $\psi^\perp=-\beta^\ast\vert0\rangle+\alpha^\ast \vert 1\rangle$
for $\psi=\alpha\vert0\rangle+\beta\vert 1\rangle$.
The most general form
of the final state is then the linear combination \be\label{ab}
A\,\psi_1\,S_{23}+B\,\psi_2\, S_{13}.
\ee
Note that $ S_{12} \psi_3 $ is not linearly independent. Eq. (\ref{ab})
can be rewritten as \be A\psi_1\left(\psi_2\psi^\perp_3-\psi^\perp_2\psi_3\right)+B\left(\psi_1\psi_2\psi^\perp_3-\psi^\perp_1\psi_2\psi_3\right)=\\
(A+B)\psi\psi\psi^\perp-\left(A\psi\psi^\perp+B\psi^\perp\psi\right)\psi.\ee
This manifestly corresponds to a universal transformation because everything
is expressed in terms of $\psi$ and $\psi^\perp$. Its linearity is obvious
by construction. The case $A=B$ is exactly the Bu\v{z}ek--Hillery symmetric
universal cloner. The cases $A\neq B$ correspond to asymmetric universal
cloners \cite{cerf,braunstein}. The third particle is an anti--clone.

As a first generalization, the above construction can be extended to
the $N\rightarrow M$ cloning case. Now linearity implies that $N$ instances
of $\psi$ have to appear in the final state, which have to be supplemented
by $M-N$ singlets. Thus, the final state is a linear combination of
the term \be \psi_1 \dots \psi_N\, S_{N+1,M+1}\dots S_{M,2M-N}\ee
and its permutations. The Gisin--Massar optimal universal cloners
correspond to the case where the above expression is symmetrized over
the first $N$ qubits. This gives an even more precise sense to our above
statement that the information about $\psi$ is spread out over the
$M$ copies in a symmetric way. Originally, the qubits $N+1,\dots,M$
contain no information because they belong to singlet states.
Not completely symmetrized linear combinations correspond to general
asymmetric cloners.

To generalize the above considerations to the $d$--dimensional case,
where there are no singlet states, one has to note that our above
definition of universality was more restrictive than the usual one.
Let us now only demand that the reduced density matrix of each clone
be of the form $s\vert\psi\rangle\langle\psi\vert+\frac{(1-s)}{d}\openone$.
Then one can replace the singlets in the above construction by
the maximally entangled state $X=\sum_{n=1}^d|n\rangle\vert n\rangle$.
Apart from this substitution the calculation remains exactly the same
as for qubits. In the $1\rightarrow 2$ case, one has the family of final
states \be A\psi_1 X_{23}+B\psi_2 X_{13}=\\
A\sum_{n=1}^d\vert\psi\rangle\vert n\rangle\vert n\rangle+
B\sum_{n=1}^d \vert n\rangle\vert \psi\rangle\vert n\rangle.
\ee
To determine the reduced density matrix of the first particle,
consider the density matrix of the first two:\be
\sum_{n=1}^d\left(  A \vert\psi\rangle\vert n\rangle +
B \vert n\rangle\vert \psi\rangle \right) \left(
A^\ast \langle \psi\vert\langle n\vert +
B^\ast  \langle n\vert\langle \psi\vert\right).\ee
Tracing over the second particle, one sees that the reduced density matrix
indeed only depends on $\vert\psi\rangle\langle \psi\vert$ and
$\openone$. Again the case $A=B$ corresponds to the general universal
optimal cloners of Bu\v{z}ek and Hillery \cite{buzekprl}. The construction
generalizes to the $N\rightarrow M$ case \cite{ddimgen}.
Thus we have derived all the optimal universal cloning transformations
in a simple and unified way.

We will come back to the above considerations
when
explaining why our stimulated emission cloners are optimal in section
\ref{whyoptimal}.
We now turn to a detailed investigation of quantum cloning by
stimulated emission. The relationship between cloning and
superluminal communication suggested by Herbert's original
proposal will be explored in the sequel.

\section{Cloning with Lambda--Atoms}\label{lambdas}

When the Bu\v{z}ek--Hillery and Gisin--Massar transformations were
discovered, the realization was first discussed in the
context of quantum computation, in other words in terms of gates.
For example, a network for the Bu\v{z}ek--Hillery cloner was
suggested in \cite{buzbraun}. However, there is actually a physical process
which seems a very natural candidate for quantum cloning, namely
stimulated emission. This was already realized by Herbert \cite{herbert},
who thought however about perfect cloning. It was then pointed out
by Mandel \cite{mandel} and Milonni and Hardies \cite{milonni} that perfect cloning is
prevented by the presence of spontaneous emission. In the new
context of non-perfect but universal cloning following Bu\v{z}ek
and Hillery, one is led to the question of how well cloning via
stimulated emission could work. Could it be optimal? We will see
that the answer is yes.

Let us first recall the basic facts about stimulated emission by
considering the simplest possible model, namely a two--level system
with levels $e$ and $g$, coupled to a single mode $a$ of the
electro--magnetic field via the Hamiltonian
\be
H=\gamma(|e\rangle\langle g|a+\mbox{h.c.}),
\ee
where $\gamma$ is a coupling constant.

This is known as the Jaynes--Cummings model. The two--level system
plays the role of inverted medium, i.e. it is supposed to be
prepared in the excited state $|e\rangle$. Now consider the transition
amplitudes for emission of a photon. For early times their size is
determined by the matrix elements of the Hamiltonian between the
respective states, i.e. by
\be \langle g|\langle n+1| H|e\rangle |n\rangle,
\ee
where we have assumed that $n$ photons are already present in node
$a$. From the commutation relation $[a,a^\dagger]=\openone$
of the bosonic operator $a$ it
follows immediately that $\langle n+1|a^\dagger|n\rangle=\sqrt{n+1}$.
Therefore this matrix element scales like
$\sqrt{n+1}$.
 This implies that the two--level system is more likely
to emit a photon in a given mode the more photons are already
present in this mode. The photons that are already present
stimulate the emission of new photons. This is of course the
fundamental principle for the working of lasers. Spontaneous
emission corresponds to the case where originally there are no
photons in mode $a$, i.e. to the matrix element $\langle e|\langle 1| H|g\rangle |0\rangle$.
Sometimes one
says that in this case the emission is stimulated by the vacuum
fluctuations.

Now imagine an inverted medium that can emit photons of different
polarizations. Two  modes, $a_1$ and $a_2$, are sufficient to describe
polarization. For the sake of clarity and simplicity we are going
to neglect the fact that in general there are different spatial
modes. If the system is in some sense rotationally invariant and
one sends in an $a_1$ photon, the emission of $a_1$ photons which is
stimulated should be more likely than the emission of $a_2$ photons,
which is only spontaneous. This implies that some kind of copying
should occur. It is also clear that this copying procedure might
not be perfect since spontaneous emission is going to happen, i.e.
in general there will be unwanted $a_2$ photons. In fact, the
presence of this spontaneous emission is unavoidable if one
demands universality of the copying, as will become clear below.
Note that if our inverted medium is really rotationally invariant
this should ensure the universality of the cloning procedure. In
such a situation spontaneous emission into both polarization modes
will be equally likely, but because of stimulation the amplitude
for emission into the desired mode will be larger, roughly by
a factor $\sqrt{n+1}$,
where $n$ is the number of incoming photons, as pointed out above.

We will start by looking at perhaps the simplest model system
which has all the required properties. It turns out that this is
already perfectly suited to achieve optimal universal cloning. The
inverted medium that we will use as a cloning device consists of
an ensemble of Lambda-atoms. These are three-level systems that
have two degenerate ground states $|g_1\rangle$ and $|g_2\rangle$
and an excited level $|e\rangle$. The ground states are coupled to
the excited state by two
 modes of the electromagnetic field, $a_1$ and $a_2$, respectively. These two modes define the
  Hilbert space of our qubit to be cloned, i.e. we want to clone general superposition
   states $(\alpha a_1^\dagger + \beta a_2^\dagger)|0,0\rangle=\alpha|1,0\rangle+\beta |0,1\rangle$.
   We can think of $a_1$ and $a_2$ as being orthogonal polarizations of one photon
   with a specific frequency, but we do not have to restrict ourselves to such a specific example,
   in fact we can think about other systems and other degrees of freedom,
   as long as they are described by the same formalism, e.g. $a_1$ and $a_2$ could also refer to
   the center-of-mass motion (phonons) in an ion trap. In the interaction
   picture, after the usual dipole and
rotating wave approximations,
   the interaction Hamiltonian between
   field and atoms has the following form:
\ba
\label{Ham1} {\cal H}_i &=& \gamma \left( a_1 \sum_{k=1}^N |e^k\rangle
\langle g^k_1| + a_2 \sum_{k=1}^N |e^k\rangle \langle g^k_2|
\right) + h.c.\nonumber \\&=&\gamma \left( a_1 \sum_{k=1}^N \sigma^k_{+,1} + a_2
\sum_{k=1}^N \sigma^k_{+,2} \right) + h.c.
\ea
The index $k$ refers to the $k$-th atom. Note that in (\ref{Ham1})
the atoms couple to only one single spatial mode of the
electromagnetic field. In particular this means that spontaneous
emission into all other modes is neglected. Situations where this
is a good approximation can now be achieved in cavity QED
\cite{flythru}. We also assume that the coupling constant $\gamma$
is the same for all atoms, which in a cavity QED setting means
that they have to be in equivalent positions relative to the
cavity mode. Trapping of atoms inside a cavity has recently been
achieved \cite{kimble}. Finally note that our Hamiltonian has no
spatial dependence, which means that the effect of the field on
the motion of the atoms is neglected, their spatial wave-function
is assumed to be unchanged. This leads to the question what the
spatial wave-function could be. The most fascinating possibility
would probably be to imagine a Bose--Einstein condensate.

The Hamiltonian (\ref{Ham1}) is invariant under simultaneous
unitary transformations of the vectors $(a_1,a_2)$ and
$(|g_1\rangle, |g_2\rangle)$ with the same matrix $U$. If one
furthermore chooses an initial state of the atoms that has the
same invariance, then the system behaves equivalently for all
incoming photon polarizations, i.e. universal cloning is achieved.
This can be seen in the following way. Consider an incident photon
in a general superposition state $(\alpha a^\dagger_1 + \beta
a^\dagger_2) |0,0\rangle$. Together with the orthogonal one-photon
state this defines a new basis in polarization space, which is
connected to the original one by a unitary transformation. If the
atomic states are now rewritten in the basis  that is connected to
the original one by the same unitary transformation, then under
the above assumptions the interaction Hamiltonian and initial
state of the atoms look exactly the same as in the original basis.
The initial state where all atoms are excited to $|e\rangle$ has
the required invariance: it is completely unaffected by the
above-mentioned transformations.

We can therefore, without loss of generality, restrict ourselves
to the cloning of photons in mode $a_1$. We consider an initial
state
\begin{equation}
|\Psi_{in}\rangle=\otimes_{k=1}^N |e^k\rangle
\frac{(a_1^\dagger)^m}{\sqrt{m!}}|0,0\rangle, \label{psii}
\end{equation}
i.e. we are starting with $m$ photons of a given polarization, and
we want to produce a certain (larger) number $n$ of clones.

\subsection{The simplest case}

For illustrative purposes let us first consider the simplest case
of one Lambda-atom and one photon polarized in direction $1$:
\begin{equation}
 |\Psi_{in}\rangle=|e\rangle
 a^\dagger_1|0,0\rangle=|e\rangle |1,0\rangle
 =: |{\cal F}_0\rangle
\end{equation}
 To
study the time development, we expand the evolution operator
$e^{-i{\cal H}t}$ into a Taylor series and determine the action of
powers of ${\cal H}$ on the state $|\Psi_{in}\rangle$.
\begin{eqnarray}
{\cal H} |\Psi_{in}\rangle&=&\gamma (|g_1\rangle a_1^\dagger
 |1,0\rangle+|g_2\rangle a_2^\dagger |1,0\rangle)\nonumber\\
 &=&\gamma \sqrt{3} \frac{  (\sqrt{2}|g_1\rangle |2,0\rangle + |g_2\rangle
 |1,1\rangle)}{\sqrt{3}}\nonumber  =: \gamma \sqrt{3}
 |{\cal F}_1\rangle \nonumber \\
{\cal H}^2 |\Psi_{in}\rangle&=&\gamma^2 (|e \rangle a_1 \sqrt{2}
 |2,0\rangle + |e\rangle a_2 |1,1\rangle)=3 \gamma^2 |e\rangle
 |1,0\rangle=3 \gamma^2  |{\cal F}_0\rangle
\nonumber \\
 & \ldots & \nonumber \\
 \label{hpowers}
 \end{eqnarray}

 The result is
 \begin{eqnarray}
 \label{onelam}
e^{-i{\cal H} t }|\Psi_{in}\rangle &=& \cos(\gamma \sqrt{3}t
)|e\rangle |1,0\rangle - i \sin{(\gamma \sqrt{3}t)}
(\sqrt{\frac{2}{3}}|g_1\rangle |2,0\rangle +\sqrt{\frac{1}{3}}
 |g_2\rangle |1,1\rangle)\nonumber\\
&=& \cos(\gamma \sqrt{3}t) |{\cal F}_0\rangle -i \sin{(\gamma
 \sqrt{3}t)} |{\cal F}_1\rangle
\end{eqnarray}
$|{\cal F}_0\rangle$ and $|{\cal F}_1\rangle$ denote the states of
the system atom-photons that lie in the subspace with $1$ and $2$
photons respectively.
 $|{\cal F}_0\rangle$ is in the subspace where no cloning has taken place and $|{\cal F}_1\rangle$
 in the one where
one additional photon has been emitted, so that the two photons
can now be viewed as clones with a certain fidelity. This way of
labeling the states will turn out to be convenient below. The
probability that the system acts as a cloner is
$p(1)=\sin^2(\gamma \sqrt{3}t)$. The fidelity $F_1$ of the cloning
procedure can be defined as the relative frequency of photons in
the correct polarization mode in the final state $|{\cal
F}_1\rangle$ (cf. Sec. \ref{comparison}). One finds
\begin{equation}
F_1=\frac{2}{3}\cdot1+\frac{1}{3}\cdot\frac{1}{2}=\frac{5}{6},
\end{equation}
which is exactly the optimal fidelity for a 1-to-2 cloner,
cf. Sec. \ref{sandc}. Actually, the state
\begin{equation}
|{\cal
F}_1\rangle=\sqrt{\frac{2}{3}}|2,0\rangle|g_1\rangle+\sqrt{\frac{1}{3}}|1,1\rangle|g_2\rangle
\label{f1}
\end{equation}
is exactly equivalent to the three-qubit state
\begin{equation}
\sqrt{\frac{2}{3}}|11\rangle|\downarrow\rangle +
\sqrt{\frac{1}{3}}\left(\frac{1}{\sqrt{2}}(|01\rangle+|10\rangle)\right)|\uparrow\rangle
\label{bh}
\end{equation}
produced by the Bu\v{z}ek-Hillery cloner.
The equivalence is established, if the photonic
states in Eq. (\ref{f1}) are identified with the corresponding
{\it symmetrized} two-qubit states (both photons in mode 1 means
both qubits in state $|1\rangle$, one photon in each mode means
one qubit in state $|1\rangle$, one in state $|0\rangle$) in Eq.
(\ref{bh}), while the atomic states $|g_1\rangle$ and
$|g_2\rangle$ are identified with the states $|\downarrow\rangle$
and $|\uparrow\rangle$ of the ancillary qubit. This is another way
of proving the optimality of Eq. (\ref{f1}). Note that in our case
the {\it universality} follows directly from the symmetry of
initial state and Hamiltonian, as explained above. In the
following we show that a similar equivalence holds between our
cloning scheme and the Gisin-Massar cloners in the completely
general case (arbitrary numbers of photons and atoms).

\subsection{Equivalence to coupled harmonic oscillators}

\label{osc}

We now turn to the discussion of the general case, i.e. we
consider the initial state (\ref{psii}). We are going to show the
equivalence of our system defined by (\ref{Ham1}) and (\ref{psii})
to a system of coupled harmonic oscillators. First note that both
the initial state (\ref{psii}) and the Hamiltonian (\ref{Ham1})
are invariant under permutations of the atoms, which implies that
the state vector of the system will always be completely
symmetric. Furthermore the Hamiltonian (\ref{Ham1}) can be
rewritten as
\begin{equation}
{\cal H}=\gamma \left( a_1 J_{+,1} + a_2 J_{+,2} \right) + h.c.
\end{equation}
in terms of ``total angular momentum'' operators
\begin{equation}
J_{+,r}=  \sum_{k=1}^N \sigma^k_{+,r} =     \sum_{k=1}^N
|e^k\rangle \langle g^k_r|           \hspace{1cm}  (r=1,2),
\end{equation}
By the above considerations one is led to use a Schwinger type
representation \cite{schwinger} for the angular momentum
operators:
\begin{equation}
J_{+,r}= b_r c^{\dagger}    \hspace{1cm} (r=1,2),
\label{schwinger}
\end{equation}
where $c^{\dagger}$ is a harmonic oscillator operator creating
``$e$'' type excitations, while $b_1$ destroys ``$g_1$''
excitations. Note that $J_{+,1}$ and $J_{+,2}$ share the operator
$c^{\dagger}$ because both ground levels $g_1$ and $g_2$ are
connected to the same upper level $e$ by the Hamiltonian
(\ref{Ham1}), and correspondingly for the Hermitian conjugates. In
terms of these operators, (\ref{Ham1}) acquires the form
\begin{equation}
{\cal H}_{osc}=\gamma ( a_1 b_1 +  a_2 b_2)c^{\dagger}+ h.c.,
\label{Hampd1}
\end{equation}
while the initial state (\ref{psii}) is now given by
\begin{equation}
|\psi_i\rangle=\frac{(a_1^{\dagger})^m}{\sqrt{m!}}\frac{(c^{\dagger})^N}{\sqrt{N!}}|0\rangle=
|m_{a1},0_{a2},0_{b1},0_{b2},N_c\rangle\equiv|m,0,0,0,N\rangle.
\label{psiinew}
\end{equation}
Actually, for reasons that will become apparent below, it is
slightly more convenient for our purposes to use the following
Hamiltonian instead of (\ref{Hampd1}):
\begin{equation}
{\cal H}=\gamma (a_1 b_2 -  a_2 b_1) c^{\dagger}+ h.c.,
\label{Hampd}
\end{equation}
which can be obtained from (\ref{Hampd1}) by a simple unitary
transformation in mode $b$, corresponding to a simple redefinition
of the atomic states in (\ref{Ham1}). This is the Hamiltonian that
is going to be used in the rest of this paper. The invariance
properties of (\ref{Hampd}) are linked to those of (\ref{Ham1}) or
equivalently (\ref{Hampd1}) discussed above: (\ref{Hampd}) is
invariant under simultaneous identical SU(2) transformations in
modes $a$ and $b$ (because the determinant of such a
transformation is equal to unity), while a phase transformation in
either mode can be absorbed into $\gamma$. This ensures the
universality of the cloning procedure.

We are now dealing with five harmonic oscillator modes defined by
the operators $c, b_1, b_2, a_1$, and $a_2$. Action of
(\ref{Hampd}) on (\ref{psiinew}) generates Fock basis states of
the general form
\begin{equation}
|(m+j)_{a1},i_{a2},i_{b1},j_{b2},(N-i-j)_c\rangle=|m+j,i\rangle_{photons}|i,j,N-i-j\rangle_{atoms}.
\end{equation}
Remember that $a_1$ is now coupled to $b_2$ etc. Expressed in
terms of individual atoms, $|i,j,N-i-j\rangle_{atoms}$ is the
completely symmetrized state with  $i$ atoms in level $g_1$, $j$
atoms in level $g_2$, and $N-i-j$ atoms in level $e$. The
correctness of (\ref{schwinger}) can be checked by explicit
application of left hand side and right hand side to such a
general state, written in terms of the individual atoms and in
terms of harmonic oscillator eigenstates respectively.

As noted above, the action of the Hamiltonian (\ref{Ham1}) on the
initial state (\ref{psii}) only generates completely symmetric
states of the atomic system. These states have the general form
\ba
\label{atomic}  {{N} \choose {i,j}}^{-1/2} \sum_{\alpha}
|g_1^{\alpha_1},g_1^{\alpha_2},\ldots,g_1^{\alpha_i},g_2^{\alpha_{i+1}},\ldots,g_2^{\alpha_{i+j}},
e^{\alpha_{i+j+1}},\ldots,e^{\alpha_N}\rangle \nonumber\\
=: | i, j, N-i-j
\rangle_{atoms}
\ea
where the sum is over all arrangements $\alpha$ of the $N-i-j$
levels $|e\rangle$, the $i$ levels $|g_1\rangle$, and the $j$
levels $|g_2\rangle$ on the $N$ atoms, and ${{N} \choose
{i,j}}=\frac{N!}{i!j!(N-i-j)!}$ is the multinomial coefficient
giving the number of such arrangements.

Now study the action of a typical term in the Hamiltonian
(\ref{Ham1}) on the system whose state we will write as\\
$|i,j,N-i-j\rangle_{atoms} \otimes |m+i,j\rangle_{photons}$:
\begin{eqnarray}
& &(\sum_{k=1}^N |g_1^k\rangle \langle e^k|) a_1^\dagger
|i,j,N-i-j\rangle_{atoms} \otimes |m+i,j\rangle_{photons}
\nonumber
\\ &=& \sum_{k=1}^N |g^k_1\rangle \langle e^k|
\sqrt{\frac{i!j!(N-i-j)!}{N!}} \sum_\alpha
|g_1^{\alpha_1},\ldots,g_1^{\alpha_i},g_2^{\alpha_{i+1}},\ldots,e^{\alpha_N}\rangle\nonumber\\
& & \otimes a_1^\dagger |m+i,j\rangle_{field} \nonumber \\ &=& (i+1)
\sqrt{\frac{i!j!(N-i-j)!}{N!}} \sum_\alpha
|g_1^{\alpha_1},\ldots,g_1^{\alpha_i},g_1^{\alpha_{i+1}},g_2^{\alpha_{i+2}},\ldots,e^{\alpha_N}\rangle \nonumber\\
& & \otimes a_1^\dagger |m+i,j\rangle_{field} \nonumber \\ &=&
\sqrt{i+1} \sqrt{N-i-j}
\sqrt{\frac{(i+1)!j!(N-i-j-1)!}{N!}}\nonumber\\
& & \sum_\alpha
|g_1^{\alpha_1},\ldots,g_1^{\alpha_i},g_1^{\alpha_{i+1}},g_2^{\alpha_{i+2}},\ldots,e^{\alpha_N}\rangle
\otimes a_1^\dagger |m+i,j\rangle_{field} \nonumber \\ &=&
\sqrt{i+1} \sqrt{N-i-j} |i+1,j,N-i-j-1\rangle_{atoms} \otimes
a_1^\dagger |m+i,j\rangle_{field}
\end{eqnarray}
Here the factor $(i+1)$ arises from the number of different
configurations that a given arrangement $\alpha$ can be reached
by. This shows that this term acts exactly like a term
$a_1^\dagger b_1^\dagger c$. Similar calculations can be made for
the other terms in the Hamiltonian. Together, they justify the
Schwinger representation (\ref{schwinger}).
 Note that the use of the Schwinger
representation is only convenient because the initial state of the
atomic system in (\ref{psii}) is completely symmetric under
permutation of the atoms.

Studying the Hamiltonian in the form (\ref{Hampd}) instead of
(\ref{Ham1}) is helpful in several respects. The number of atoms
$N$ that is explicit in the Hamiltonian (\ref{Ham1}) now appears
only as a part of the initial conditions of our system, which
makes it easy to treat the general case of N atoms in one go. We
will do this in the next subsection.

The Hamiltonian (\ref{Hampd}) can also be seen as a Hamiltonian
for down-conversion with a quantized pump-mode described by the
operator $c$, while $a_r$ and $b_r$ are the signal and idler modes
respectively, where $r$ labels the polarization degree of freedom.
Usually in parametric down--conversion
the operator $c$ of (\ref{Hampd}) is replaced by a
c--number.  This corresponds to
the limit of a classical pump field.
These remarks lead to an experimental realization of optimal quantum
cloning via stimulated emission which we will discuss in Sec. \ref{expreal}.

In passing we note that the above dynamical equivalence generalizes
to atoms with more than $2$ ground-states $|g_n\rangle$ that are
coupled each to a different degree of freedom of photons $a_n$. By
similar arguments a system of $N$ identical atoms with $r$ ground
states $\{|g_1\rangle,\ldots,|g_r\rangle\}$ governed by a
Hamiltonian
\begin{equation}
{\cal H}^r=\gamma \sum_{k=1}^N \sum_{n=1}^r |e^k\rangle \langle
g_n^k| a_n +h.c.
\end{equation}
is equivalent to a system of $r+1$ coupled harmonic oscillators with
lowering operators $c$ and $b_1, \ldots , b_r$ governed by the interaction Hamiltonian
\begin{equation}
{\cal H}^r_{osc}=\gamma \sum_{n=1}^r cb_n^\dagger a_n^\dagger +
h.c.
\end{equation}

\subsection{Cloning of $m$ photons with $N$ Lambda-atoms: Proof of optimality}

\label{general}

We are now going to show that the system defined by
(\ref{psiinew}) and (\ref{Hampd}) indeed realizes optimal cloning
for arbitrary $N$ and $m$. The idea of the proof is the following.
After evolution in time the system that started with a certain
photon number $m$ will be in a superposition of states with
different total photon numbers, where total means counting photons
in mode $a_1$ and $a_2$, i.e. both ``good'' and ``bad'' copies. We
will show that the general form of the state vector after a time
interval $t$ is
\begin{equation}\label{psif}
|\Psi(t)\rangle = e^{-i {\cal H} t}
 |\Psi_{in}\rangle =\sum_{l=0}^N f_l(t) |{\cal F}_l\rangle,
\end{equation}
where $l$ denotes the number of {\it additional} photons that have
been emitted and
\be
\label{Fs} |{\cal F}_l\rangle:=\\{{m+l+1}\choose {l}}^{-\frac{1}{2}}
\sum_{i=0}^l (-1)^i \sqrt{{m+l-i} \choose {m}}
 |(m+l-i)_{a1},i_{a2},i_{b1},(l-i)_{b2},(N-l)_c\rangle.
\ee
Note that the number of photons can never become smaller than $m$
since all the atoms start out in the excited state. $|{\cal F}
_l\rangle$ is a normalized state of the system with $m+l$ photons
in total. To see that $|{\cal F}_l\rangle$ is properly normalized
note that $\sum_{i=0}^l {{m+i} \choose {m}} = {{m+l+1} \choose
{l}}$.

The states $|{\cal F}_l\rangle$ are formally identical to the
states obtained in \cite{buzeknot}, which have been shown to realize
optimal universal cloning and the optimal universal NOT
simultaneously. The ideal universal NOT is an operation that
produces the orthogonal complement of an arbitrary qubit. Like
perfect cloning, it is prohibited by quantum mechanics. The
transformation in \cite{buzeknot} links universal cloning and
universal NOT (anti-cloning): the ancilla qubits of the cloning
transformation are the anti-clones. In our case, the clones are
the photons in the $a$-modes and the anti-clones are the atoms in
the $b$-modes (atomic ground states $g_1$ and $g_2$). From the
Hamiltonian (\ref{Hampd}) and (\ref{Fs}) it is clear that for
every ``good'' emitted photon-clone (in mode $a_1$) there is an
excitation in mode $b_2$ which corresponds to an anti-clone
(atomic ground state $|g_2\rangle$). The only difference to the
states in \cite{buzeknot} is the presence of the fifth harmonic
oscillator mode $c$, describing the ``e'' type excitations, which
counts the total number of clones that have been produced (equal
to the number of atoms having gone to one of the ground states)
and doesn't affect any of the conclusions.

A distinguishing feature of our cloner is that the output state
(\ref{Fs}) is a superposition of states with different total
numbers of clones. Cloning with a certain fixed number of produced
copies can be realized by measuring the number of atoms in the
excited state $|e\rangle$ (corresponding to mode $c$) and
post-selection.

To prove that the system is indeed always in a superposition of
the states $|{\cal F}_l\rangle$ as in Eq. (\ref{Fs}) we use
induction: The initial state of the system is
$|\Psi_{in}\rangle=|{\cal F}_0\rangle$. Now we will show that if
$|\Phi\rangle$ is a superposition of states $|{\cal F}_l\rangle$
then ${\cal H} |\Phi\rangle$ is so, too. Then, since
$|\Psi(t)\rangle=e^{-i {\cal H} t} |\Psi_{in}\rangle=\sum_{p}
\frac{(-i{\cal H} t)^p}{p!} |\Psi_{in}\rangle$ this implies that
$|\Psi(t)\rangle$ will be a superposition of $|{\cal F}_l\rangle$.
Explicit calculation shows that
\begin{eqnarray}
{\cal H} |{\cal F}_l\rangle & = & \gamma
(\sqrt{(l+1)(N-l)(m+l+2)}|{\cal F}_{l+1}\rangle\nonumber\\
& &+\sqrt{l(N-l+1)(m+l+1)}  |{\cal F}_{l-1}\rangle) \quad 1 \leq l < N
  \nonumber \\{\cal H} |{\cal F}_0\rangle & = & \gamma \sqrt{N(m+2)}|{\cal F}_{1}\rangle \nonumber \\
{\cal H} |{\cal F}_N\rangle & = & \gamma \sqrt{N(m+N+1)} |{\cal
F}_{N-1}\rangle \label{rec}
\end{eqnarray}
which completes the proof.

Note that the form of the coefficients $f_{l}(t)$ didn't play any
role in our proof. Actually, the $f_l$ are in general hard to
determine exactly. Solutions have been found in limiting cases.
For the limit of a classical pump field ($c$ replaced by a
c--number), the solution can be found by standard methods and is
given in Sec. \ref{expreal}
in the context of a proposed experimental realization of quantum cloning.
The solution in the case of large incoming
photon numbers ($m\gg N$) can be obtained in the following way.

For that case, the recursion (\ref{rec}) becomes
\begin{eqnarray}
{\cal H} |{\cal F}_l\rangle & = & \gamma
\sqrt{m}(\sqrt{(l+1)(N-l)}|{\cal
  F}_{l+1}\rangle+  \sqrt{l(N-l+1)}  |{\cal F}_{l-1}\rangle) \quad 1 \leq
  l < N
  \nonumber \\{\cal H} |{\cal F}_0\rangle & = & \gamma \sqrt{m}\sqrt{N}|{\cal
  F}_{1}\rangle \nonumber \\
{\cal H} |{\cal F}_N\rangle & = & \gamma \sqrt{m}\sqrt{N} |{\cal
F}_{N-1}\rangle \label{recm}
\end{eqnarray}
It is possible to diagonalize the ``transfer'' matrix $A$ acting on the vector
$(f_0,\ldots,f_N)$ that corresponds to the action of ${\cal H}$ on
$|\Psi\rangle=\sum_{l=0}^N f_l |{\cal F}_l \rangle$:
\be
A_{l,l+1}=\gamma \sqrt{m}
\sqrt{(l+1)(N-l)} =A_{l+1,l}. \ee
This allows to exponentiate $A$ and to determine
the final state of the system after a time $t$: \begin{equation} \label{soln}
|\Psi(t)\rangle=\sum_{l=0}^N (-i)^l \sqrt{{{N} \choose {l}}} \cos^{N-l}( \gamma
\sqrt{m} t) \sin^l( \gamma \sqrt{m} t) |{\cal F}_l\rangle \end{equation}
Differentiating (\ref{soln}) and using (\ref{recm}) one can show that this
state fulfills Schr\"{o}dinger's equation with the correct initial
condition.

In this big-$m$-limit the probability to observe the system as an
$m\rightarrow m+l$ cloner (i.e. the probability that $l$
additional photons are emitted) is
\begin{equation}
p(l)={{N} \choose {l}} \cos^{2(N-l)}(\gamma \sqrt{m} t) \sin^{2l}(
\gamma \sqrt{m} t)
\end{equation}
This is a binomial distribution with a probability $\sin^2 (\gamma
\sqrt{m} t)$ for each atom to emit a photon. Setting $N=1$ or
comparison with Eq. (\ref{onelam}) shows that this is identical to
the probability for the case of only one atom in the case of large
$m$. This means that in this limit each atom interacts
independently with the electromagnetic field, because the effect
of the other atoms on the field is negligible. In the short-time
limit $p(l)=O(t^{2l})$. Furthermore the expected average number of
``clones'' $N_c=\sum_{l=0}^N l p(l)=N \sin^2(\gamma \sqrt{m} t)$
oscillates with an $m$-dependent frequency.

Let us pause here for a moment and summarize what we have found.
Our system consisting of an ensemble of Lambda-atoms in the
excited state is indeed equivalent to a superposition of optimal
cloning machines a la Bu\v{z}ek-Hillery or Gisin-Massar, producing
various numbers of clones. The atoms play the double role of
photon source and of ancilla, the atomic ground states can be
identified with the ancilla states in the qubit cloners. As for
the corresponding qubit cloners, those ancillary atoms can also be
seen as the output of a universal NOT gate. On the other hand, the
atoms that end up in the excited state provide information about
the number of clones that has actually been produced. This can be
used to realize cloning and anti--cloning with a fixed number of output clones by
post--selection.

\subsection{The equivalence between pairs of V--atoms and Lambda--atoms}
\label{vs}

In this section we present an alternative (but similar) way of
realizing optimal universal cloning that uses entangled pairs of
V-atoms instead of Lambda atoms. We prove optimality by showing
that the system can be exactly mapped onto the system with Lambda
atoms that we discussed above.

The two degenerate upper levels of each V-atom, $|e_1\rangle$ and
$|e_2\rangle$, are coupled to the ground state $|g\rangle$ via the
two orthogonal
 modes $a_1$ and $a_2$ respectively. The Hamiltonian describing the interaction of atom and field is:
\ba
\label{Ham2} {\cal H}_V&=& \gamma \left(a^\dagger_1 \sum_{k=1}^N
|g^k\rangle \langle e^k_1| +a^\dagger_2 \sum_{k=1}^N |g^k\rangle
\langle e^k_2| \right) + \mbox{h.c.}\nonumber\\
&=& \gamma \left(a^\dagger_1
\sum_{k=1}^N \sigma^k_{-,1} +a^\dagger_2 \sum_{k=1}^N
\sigma^k_{-,2} \right) + \mbox{h.c.}
\ea
It arises from similar assumptions as (\ref{Ham1}). In contrast to
before we now choose an entangled state of the atoms as the
initial state. This is motivated by the fact that the initial
atomic state has to be a {\it singlet} under polarization
transformations in order for our cloning device to be again
universal.

Let us first examine the simplest case of two entangled V-atoms,
$A$ and $B$, and one incoming photon. The initial state of the
system is
\begin{equation}
|\Psi_{in}\rangle=\frac{1}{\sqrt{2}}(|e_1^A e_2^B\rangle-|e_2^A
 e_1^B\rangle)\otimes |1,0\rangle
\end{equation}
Developing the time evolution operator $e^{-i{\cal H_V}t}$ into a
power series, one finds easily:
\begin{eqnarray}
& e^{-i{\cal H}_V t }|\Psi_{in}\rangle = \cos(\gamma \sqrt{3}t)
\frac{|e_1^A e_2^B\rangle - |e_2^A e_1^B\rangle}
{\sqrt{2}}|1 ,0\rangle &\nonumber \\
& - i \sin{(\gamma \sqrt{3}t)}
\left(\sqrt{\frac{2}{3}} \frac{|g^A
e_2^B\rangle - |e_2^A g^B\rangle}{\sqrt{2}} |2,0\rangle
+\sqrt{\frac{1}{3}} \frac{|e_1^A g^B\rangle - |g^A
e_1^B\rangle}{\sqrt{2}} |1,1\rangle \right) \label{twov}&
\end{eqnarray}
With the substitution
\begin{eqnarray}
\label{subst} \frac{|e_1^A e_2^B\rangle-|e_2^A
e_1^B\rangle}{\sqrt{2}} \longrightarrow |\tilde{e}\rangle
\nonumber\\ \frac{|g^A e_2^B\rangle-|e_2^A g^B\rangle}{\sqrt{2}}
\longrightarrow |\tilde{g_1}\rangle \nonumber\\ \frac{|e_1^A
g^B\rangle-|g^A e_1^B\rangle}{\sqrt{2}} \longrightarrow
|\tilde{g_2}\rangle
\end{eqnarray}
the state (\ref{twov}) has exactly the same form as the
corresponding state (\ref{onelam}) for one Lambda-atom, which
implies that it also implements optimal universal $1\rightarrow 2$
cloning.

Actually, the correspondence goes much further. Consider an
initial atomic state consisting of $N$ pairs of V-atoms, where
each pair is in a singlet state:
\begin{equation}
|\psi_i\rangle=\otimes_{k=1}^N |\tilde
  e^k\rangle
\end{equation}
with $|\tilde e\rangle$ as defined in (\ref{subst}).

It is easy to see that the action of the Hamiltonian (\ref{Ham2})
on each pair only
  generates one of the three antisymmetric atomic states in Eq. (\ref{subst}). Because of the
  invariance of the Hamiltonian under permutations, and in particular under the exchange
  of two atoms belonging to the same pair, transitions between states with different symmetry
  properties are impossible.
  In fact, with the identification (\ref{subst}) the Hamiltonian (\ref{Ham2}) has exactly the same form
  as the
  Hamiltonian for Lambda-atoms (\ref{Ham1}).
The analysis made previously for Lambda atoms now
goes through unchanged and we obtain the same cloning properties
of a system of pairwise entangled V-atoms as we had before for
Lambda-atoms, i.e. we have found another way of realizing optimal
universal cloning. Although this scheme would without doubt be
more difficult to realize experimentally, we believe that the
underlying equivalence between the two systems is interesting and
may be useful in other contexts as well.

\subsection{Single V--Atoms are sub-optimal cloners}\label{subop}

From the results of the previous sections, one might be tempted
to conclude that the fulfillment of the symmetry requirements
discussed above already implies optimality of the cloning procedure.
Here we show that this is not the case by studying an explicit example
of universal but suboptimal cloning via stimulated emission.

We are again considering an ensemble of V--atoms where
each atom is initially in the mixed state
\begin{equation}
\rho_{i}=\frac{1}{2} (|e_1\rangle \langle e_1| +|e_2\rangle
\langle e_2| ),
\end{equation}
which is invariant under the same unitary transformations. The
invariance of both Hamiltonian and initial state together ensure
the universality of the cloning procedure. Therefore it is
sufficient to analyze the performance of the cloner for one
arbitrary incoming one-photon state; we choose
$|\psi_i\rangle=a^\dagger_1 |0\rangle$.

We have performed numerical computations for
systems of a few (up to $N=6$) atoms.  From (\ref{Ham2}), the
time development operator $U=e^{-iHt}$ for the whole atoms-photons
system was calculated. Use was made of the fact that $N_1$ and
$N_2$, which denote the sum of the number of photons plus the
number of excited atoms for mode 1 and 2 respectively, are
independently conserved quantities. Therefore the whole Hilbert
space is decomposable into invariant subspaces, i.e. $H$ and $U$
are block-diagonal.

The final state of the procedure is an entangled state of the
atom-photon system that has components with various numbers of
photons, where the maximum possible total number is $N+1$ (if all
atoms have emitted their photons). The probability to find $k$
``right'' and $l$ ``wrong'' photons in the final state, denoted by
$p(k,l)$, was calculated for all possible values of $k$ and $l$
and for different values of $\gamma t$, and from it the overall
average ``fidelity''

\begin{equation}
f_{clones}(t)\!=\! \sum \limits_{k+l\geq2} p'(k,l;t) \left(
\frac{k}{k+l} \right) \label{fclones}
\end{equation}
was determined. This is the average of the relative frequency of
photons with the correct polarization in the final state. Note
that in (\ref{fclones}) the average is performed only over those
cases where there are at least two photons in the final state,
i.e. where at least one clone has been produced.
$p'(k,l)=p(k,l)/(1-p(1,0)-p(0,1))$ is used in order to have proper
normalization. Note that $p(0,0)$ is always zero.

That average fidelity for our cloning procedure was compared to
the average fidelity that would be achieved by
 an ensemble of optimal
cloners producing the same distribution of numbers of photons,
i.e. to
\begin{equation}
f_{opt}(t)\!=\! \sum \limits_{n=2}^{N+1} p'(n;t)\left( \frac{2n+1}{3n}
\right),\label{fopt}
\end{equation}
where $p'(n)\!=\!\sum \limits_{k+l=n} p'(k,l)$. We also made a
comparison to the case, where, in addition to the incoming photon,
photons are just created randomly, i.e. to the fidelity
\begin{equation}
f_{rand}(t)\!=\!\sum \limits_{n=2}^{N+1} p'(n;t) \left( \frac{n+1}{2n}
\right).\label{frand}
\end{equation}

\begin{figure}

        \includegraphics[width=  0.8 \columnwidth] {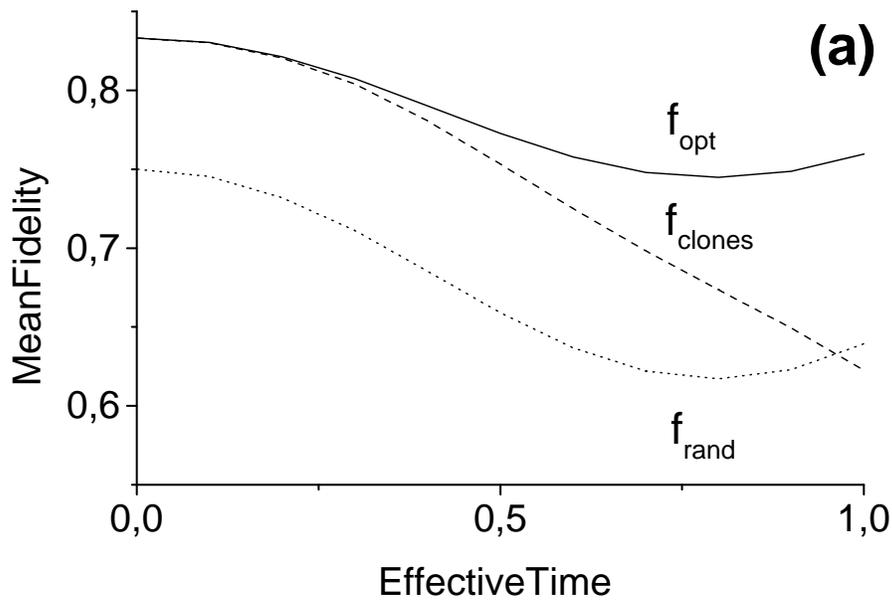}
        \caption{Dependence on time, measured in units of $\gamma t$, of $f_{opt}$, $f_{clones}$,
         and $f_{rand}$, which are the optimum possible fidelity, the
         fidelity achieved by our V--atom cloning procedure, and the fidelity achieved by random photon
        production respectively, as defined in Eqs. (\ref{fopt},\ref{fclones},\ref{frand}),
for the case of $N=6$ atoms.
        It is evident that optimal cloning is achieved in the short-time
        limit. The behavior for lower atom numbers is the same.
        }
        \label{fids}
\end{figure}

\begin{figure}
       \includegraphics[width=  0.8 \columnwidth] {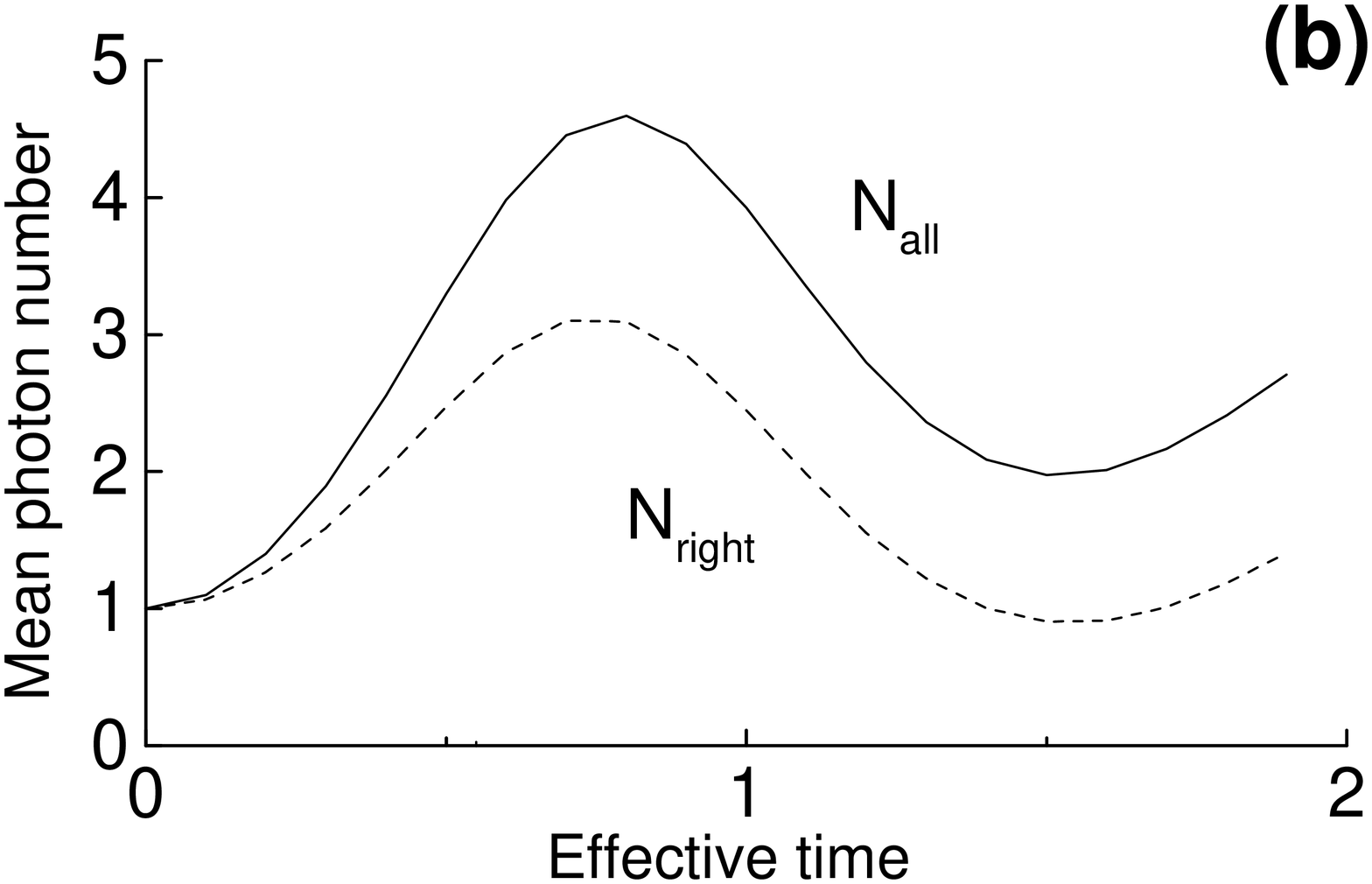}
        \caption{
 Time dependence of the mean number
        of all photons $N_{all}$ and of the mean number of ``right''
        photons (i.e. of the same polarization as the incoming
        photon)
        $N_{right}$ for the case $N=6$.
        }
        \label{photons}
\end{figure}

Fig. \ref{fids} shows clearly that the fidelity of our cloning
procedure approaches the optimum fidelity for early times. One can
also see that for longer interaction times $f_{clones}$ departs
from $f_{opt}$ and even becomes lower than $f_{rand}$. This
behavior, which may seem surprising, is due to the fact that for
longer times absorption of photons by atoms that have already
emitted once and gone to the ground-state becomes important.
 Note that absorption of ``right'' photons
is favored if there are more such photons present. In particular,
also the incoming ``right'' photon can be absorbed by an atom that
has emitted a ``wrong'' photon before, resulting in departure from
optimality for later times. The superiority of $f_{rand}$ in that
regime is understandable because in our idealized random cloner
the incoming photon is always left intact.

Our computations show that the system goes through many
emission-reabsorption cycles, though without exhibiting a simple
periodicity. As a consequence, over long times $f_{clones}$
oscillates taking values above and below $f_{rand}$, sometimes
approaching $f_{opt}$ again.

Fig. \ref{photons}, which also illustrates the above-mentioned cyclic
behavior of our system, shows the time dependence of the mean
number of photons and of the mean number of photons of the correct
polarization. For short times, which is the interesting regime
from the point of view of cloning, the probability for every
individual atom to have already emitted its photon is low.
Therefore, in order to produce a reasonable average number of
clones in this regime, a large number of atoms is necessary.

Our results show that symmetry alone is not sufficient to achieve
optimal cloning. But note that even in this case optimality is approached
for short interaction times.

\section{Experimental realization}\label{expreal}

\begin{figure}
\centering
\includegraphics[width=\textwidth]{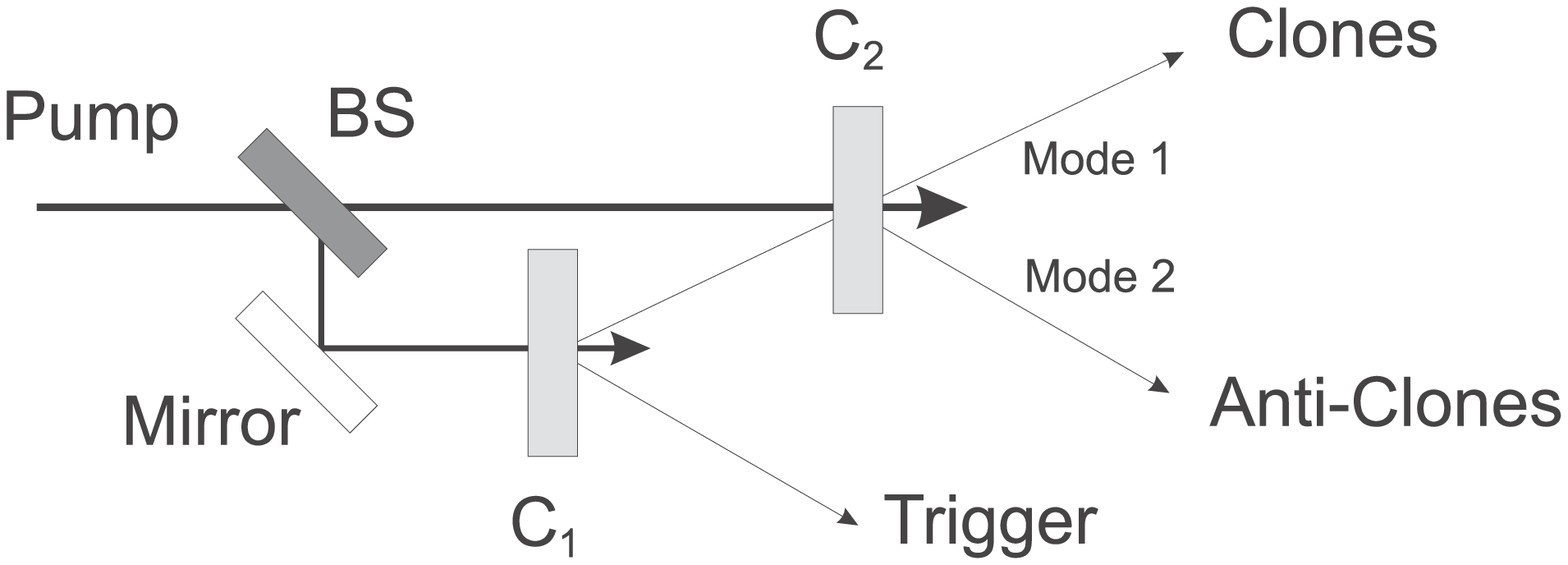}
\caption{Setup for optimal cloning by parametric down-conversion
        { \protect \cite{mandel,DeM,Migdall}}.
The pump-pulse is split at the beam splitter BS. One part of the
pump pulse hits the first crystal C$_1$, where photon pairs are
created with a certain rate.
 One photon from  each pair can be used as a trigger.  The other photon is
the system to be cloned. This photon is directed towards the
second crystal C$_2$, where it stimulates emission of photons of
the same polarization along the same direction. The path lengths
have to be adjusted in such a way that the DC-photon and the
second part of the pump pulse reach C$_2$ simultaneously. The
photons in mode 1 are optimal clones of the incoming photon, and
the photons in mode 2 are the output of an optimal universal
NOT-gate. It is interesting to note that in this scheme one is
actually cloning a photon that is part of an entangled pair. }
\label{pdc}
\end{figure}

Here we propose a concrete experimental realization of the ideas
discussed in the previous sections. The
scheme for quantum cloning that we want to present is
based on stimulated parametric down-conversion (PDC). We will show
that optimal cloning can be realized with present technology.
In PDC a strong light beam
is sent through a crystal. There is a certain (very low)
probability for a photon from the beam to decay into two photons
such that energy and crystal momentum are conserved. In type-II
PDC the two photons that are created have different polarization.
They are denoted as signal and idler.

Fig. \ref{pdc} shows the setup that we have in mind. We consider pulsed
type-II  frequency-degenerate PDC. It is possible to choose two
conjugate directions for the signal and idler beams such that
photon pairs that are created along these two directions are
entangled in polarization \cite{Source}. We consider the
quasi-collinear case (i.e. the two directions almost coincide), so
that the transverse motion of the photons in the crystal is not
important.

For stimulated emission to work optimally, there has to be maximum
overlap of the amplitudes of the incoming photon and of all the
photons that are produced in the second crystal. This can be
achieved by using a pulsed scheme together with filtering of the
photons before detection \cite{ghzvorschlag}. The pump pulse can
be seen as an active volume that moves through the crystal. If the
photons are filtered so much that the smallest possible size of
the wavepackets detected is substantially bigger than the pump
pulse, then there is maximum overlap between different pairs
created in the same pulse.  Of course, filtering limits the
achievable count rates. Moreover the group velocities of pump
pulse, signal ($V$) and idler ($H$) photons are not all identical.
This leads to separations (of the order of a few hundred fs per
millimeter in BBO), which have to be kept small compared to the
size of the DC-photon wave packets. There is a trade-off between
filtering and crystal length, i.e. one can choose narrower filters
in order to be able to use a longer crystal (which leads to longer
interaction times).

If the above-mentioned conditions are fulfilled, then a single
spatial mode (i.e. one mode for the signal and one for the idler
photons) approximation can be used. The PDC process can then be
described in the limit of a large classical pump pulse, in the
interaction picture, by the Hamiltonian
\begin{equation}
H=\gamma(a^\dagger_{V1} a^\dagger_{H2} - a^\dagger_{H1}
a^\dagger_{V2}) + h.c.,
\end{equation}
where $a^\dagger_{V1}$ is the creation operator for a photon with
polarization V propagating along direction 1 etc. The coupling
constant and the intensity of the classical pump pulse are
contained in $\gamma$. As discussed above this Hamiltonian
corresponds to the limit of the $\Lambda$--atom Hamiltonian
(\ref{Ham1}) for a coherent state of mode $c$.

The Hamiltonian $H$ is invariant under simultaneous general
$SU(2)$ transformations of the polarization vectors $(a^\dagger_V,
a^\dagger_H)$ for modes $1$ and $2$, while a phase transformation
will only change the phase of $\gamma$. This makes our cloner
universal, i.e. its performance is polarization independent.
 Therefore it is again
sufficient to analyze the ``cloning'' process in one basis.

The time development operator $e^{-iHt}$ clearly factorizes into a
$V1-H2$ and an $H1-V2$ part. Consider cloning starting from $N$
identical photons in the initial state
$|\psi_i\rangle=\frac{(a^\dagger_{V1})^N}{\sqrt{N!}} |0\rangle$
Making use of the disentangling theorem \cite{Walls,Collett} one finds
that (cf. \cite{DeM})
\begin{eqnarray}
|\psi_f\rangle=e^{-iHt}|\psi_i\rangle=K& & \sum
\limits_{k=0}^{\infty}(-i \Gamma)^k\sqrt{{k+N \choose N}
}|k+N\rangle_{V1}|k\rangle_{H2}\nonumber\\ \times &
&\sum\limits_{l=0}^{\infty} (i
\Gamma)^l|l\rangle_{H1}|l\rangle_{V2} \label{finalstate}
\end{eqnarray}
where $\Gamma=\tanh \gamma t$ and $K$ is a normalizing factor.
This reproduces the results of the previous section
c.f. Eq. (\ref{psif},\ref{Fs}). We see, that in this limiting case it is possible
to determine the coefficients $f_l(t)$ of Eq. (\ref{psif}) explicitly.

The component of this state which has a fixed number $M$ of
photons in mode 1, is proportional to
\begin{equation}
\sum \limits_{l=0}^{M-N} (-1)^l \sqrt{{M-l \choose N}}
|M-l\rangle_{V1} |l\rangle_{H1} |l\rangle_{V2} |M-N-l\rangle_{H2}.
\label{subspacestate}
\end{equation}
This is identical to the state produced by the unitary
transformation written down in \cite{buzeknot} which can be seen as a
special version of the Gisin-Massar cloners \cite{gisinmassar}
that implements optimal universal cloning and the optimal
universal NOT-gate at the same time. The $M$ photons in mode 1 are
the clones, while the $M-N$ photons in mode 2, which act as
ancillas for the cloning, are the output of the universal
NOT-gate, the ``anti-clones''.

This means that the setup of Fig. \ref{pdc} works as an ensemble
of optimal universal cloning (and universal NOT) machines,
producing different numbers of clones and anti-clones with certain
probabilities. Note that each of the modes can be used as a
trigger for the other one and therefore cloning or anti-cloning
with a fixed number of output-systems can be realized by
post-selection.

We have shown a method of realizing optimal quantum cloning
machines. We emphasize that this scheme should be experimentally
feasible with current technology. In our group, pair production
probabilities of the order of $4 \cdot 10^{-3}$ have been achieved
with a 76 MHz pulsed laser system (UV-power about 0,3 W) and a 1
mm BBO crystal, for 5 nm filter bandwidth. Past experiments show
that good overlap of photons originating from different pairs is
achieved under these conditions. With detection efficiencies
around 10 percent, this leads to a rate of two-pair detections of
the order of one per a few seconds.

Let us note that experiments in the spirit of the present proposal
are currently under way in at least two laboratories \cite{DeM, Dik}.
First results were reported in \cite{mussi}.

\section{Cloning of photons versus cloning of qubits}

\label{comparison}

In this section we are going to discuss the physical differences
that exist in spite of the formal equivalence proven above between
our photon cloners based on stimulated emission and the qubit
cloners as usually considered \cite{buzekhillery,gisinmassar}. In
particular, we will show that the claim that optimal cloning is
realized by our devices is justified in spite of these
differences.

In most of the previous work cloning was discussed in terms of
quantum networks. In general, the situation considered in these
papers is the following: one has a certain number of qubits that
are localized in different positions, which makes them perfectly
distinguishable. At the beginning, some of those qubits are the
systems to be cloned, the others play the role of ancillas. After
the cloning procedure, which consists of several joint operations
on the qubits that can be expressed in terms of quantum gates,
some of the qubits are the clones, the rest are ancillas, which
for a specific form of the optimal cloning transformation can also
be seen as outputs of the universal NOT operation. As a
consequence of localization, it is possible to address individual
clones.

In our stimulated emission cloners, the situation is different.
All input systems (photons) are in the same spatial mode (called
mode $a$), and, even more importantly, all clones
are produced into that mode. Note that this is completely
unavoidable if stimulated emission is to be used. One can say that
this is the price one has to pay for the great conceptual
simplicity of the cloning procedure itself.

However, having all clones in the same spatial mode is not
necessarily an important disadvantage. For example, if perfect
cloning of that kind were possible, one could still determine the
polarization of the original photon to arbitrary precision by
performing measurements on the clones. This would still make
superluminal communication possible.
It may be
interesting to note that in the paper that started the whole
discussion about quantum cloning, Herbert \cite{herbert}
considered cloning via stimulated emission and therefore
necessarily into a single spatial mode.

If one wants to
distribute the clones to different locations, this can for example
be achieved using an array of beam splitters. However, this does
not lead to a situation where one can be sure to have exactly one
photon in each mode. If one wants to have at most one photon in
each mode, the array has to have many more output modes than there
are photons.

Another distinguishing feature of our cloners compared to the
usual qubit cloners is the fact that the same procedure is used to
produce different numbers of clones. While in the qubit case the
network to be used depends on the number of desired clones, in our
case the final state is a superposition of states with different
numbers of clones. Of course, the average number of clones
produced depends on the number of atoms present in the system and
the interaction time. As discussed in Sec. \ref{lambdas} cloning
with a fixed number of output clones can be achieved by
post--selection based on a measurement of the number of excited
atoms in the final state.

The formal equivalence between the qubit cloners and our one-mode
cloners can arise because the output state produced by the optimal
qubit cloners is completely symmetric under the exchange of clones
\cite{buzekhillery,gisinmassar}. Because of the bosonic nature of
the photons there is a one-to-one-mapping between completely
symmetric qubit states and photonic states. Note that
asymmetric cloning could not be realized by the presented method.
For a completely
symmetric qubit state the two concepts of relative frequency of
qubits in the ``correct'' basis state and of single-particle
fidelity are equivalent. This can be seen in the following way.
Let $|\psi\rangle$ denote the state that is to be copied. Then the
usual definition of the (single--particle) cloning fidelity is
\begin{equation}
F=\langle\psi|\rho_{red} |\psi\rangle,
\end{equation}
where $\rho_{red}$ is the reduced density matrix of one of the
clones, say the first one, i.e.
\begin{equation}
\rho_{red}=\mbox{Tr}_{2,3,...,N}\left[\, \rho \, \right]
\end{equation}
 Then $F$ can also be expressed as
\begin{equation}
F=\mbox{Tr}\left[\, \rho \; |\psi\rangle\langle\psi|_1\otimes
I_2\otimes...\otimes I_N\right]. \label{F}
\end{equation}
On the other hand, the relative frequency of qubits in the state
$|\psi\rangle$ can be written as
\begin{equation}
\frac{1}{N}\mbox{Tr}\left[\, \rho \,
\left(|\psi\rangle\langle\psi|_1\otimes I_2\otimes...\otimes I_N +
I_1 \otimes |\psi\rangle\langle\psi|_2 \otimes ... \otimes I_N +
... + I_1 \otimes ... \otimes |\psi\rangle\langle\psi|_N \right)
\right]. \label{rel}
\end{equation}
If $\rho$ is invariant under exchange of any two clones, it is
obvious that (\ref{rel}) is equal to (\ref{F}), i.e. for symmetric
cloners the two concepts are completely equivalent. This justifies
our definition of fidelity via the relative frequency in the case
of photon cloning (cf. Sec. \ref{lambdas}).

Let us finally address the issue of optimality in the context of
stimulated emission cloners. In this paper we have shown the
formal equivalence of our scheme and the optimal schemes for qubit
cloning. As a consequence, the fidelity of the clones saturates
the bounds derived for the cloning of qubits. However, it is not
entirely obvious that the bounds derived for the situation of
distinct well-localized qubits also apply to our situation. Could
one maybe achieve even higher fidelity in our one-mode case? The
following argument shows that the bounds indeed apply in our
situation as well, i.e. that photon cloning is not allowed to be
better than qubit cloning.

Let us assume that we had a single-mode cloning machine that
clones photons with a better fidelity than given by the bounds for
qubits. Consequently, the relative frequency of ``correct''
photons has to exceed the bound for at least one value of the
final total photon number $M$. This is obvious if $M$ has been
fixed by post-selection. Otherwise the fidelity has to be defined
as the average of the relative frequencies over all final total
photon numbers. This average can only exceed the bound for qubits
if the bound is violated for at least one particular value $M$ of
the final photon number.

As a consequence, we have a universal map from the $N$-photon
Hilbert space to the $M$-photon Hilbert space that achieves a
relative frequency of correct photons in the final state that is
higher than the qubit bound. But the existence of such a map is
equivalent to the existence of a universal map from the totally
symmetric $N$-qubit space to the totally symmetric $M$-qubit space
with a single-particle fidelity equal to the relative frequency.
The existence of the latter map is excluded by the theorems on
cloning of qubits \cite{werner}. This justifies our claim that the
schemes presented in the previous sections realize {\it optimal}
cloning of photons.

\section{Why are our cloners optimal?}\label{whyoptimal}

On the previous pages we have shown that optimal cloning can
indeed be realized by stimulated emission. The states produced by
our simple quantum optical model systems consisting of two
photonic modes and an ensemble of Lambda systems are exactly the
same as those derived by Bu\v{z}ek and Hillery and Gisin and
Massar. But why is that so? An element of wonder seems to remain.
It follows from symmetry considerations that our systems should
act as universal cloners, but a priori optimality was not
necessarily to be expected. We have seen in subsection \ref{subop} that not
all systems which have the required symmetries also lead to
optimal universal cloning.

The reasons behind the optimality of our cloning procedure can be
understood by remembering
our construction of the optimal universal cloning transformations
in section \ref{trafo}, where we saw that the output of the
optimal cloner is given by the projection of the state
\be\label{optclone}
\psi_1 \dots \psi_N\, S_{N+1,M+1}\dots S_{M,2M-N}\ee onto the completely
symmetric subspace of the first $M$ qubits, where $S$ can be replaced
by the maximally entangled state $\frac{1}{\sqrt{2}}(|0\rangle
|0\rangle+|1\rangle|1\rangle)$.

On the other hand, one can show that the final states of
our stimulated emission cloners are always linear combinations of
states of the following form:
\be\label{phcloner}
 (a^\dagger_1 b^\dagger_1+a^\dagger_2 b^\dagger_2)^{M-N}{a^\dagger_1}^N|0\rangle,\ee
where the $c$ modes are disregarded because they just count the
number of photons, as explained above. Here an initial state with
$N$ photons in mode 1 was assumed and we have already chosen the
exponent $M-N$ in such a way as to facilitate the comparison
to (\ref{optclone}). That the final state can indeed be expressed in terms of
(\ref{phcloner}) can be seen by noting that all terms generated during the time
evolution will be of the form
\be H^k {a_1^\dagger}^N\vert 0\rangle\ee
for some $k$,
where
\be H=\gamma c(a_1^\dagger b_1^\dagger +a_2^\dagger b_2^\dagger  )+h.c.\ee

Furthermore
the commutator
\be [a_1 b_1+a_2 b_2,a_1^\dagger b_1^\dagger +a_2^\dagger b_2^\dagger  ]=
N_a+N_b+\openone,\ee
where $N_a=a_1^\dagger a_1 +a_2^\dagger a_2$ etc.
This means that we can get rid of all annihilation operators by
commuting them to the right.

The similarity between the two expressions (\ref{optclone})
and (\ref{phcloner}) is obvious.
The ${a_1^\dagger}^N$ in (\ref{phcloner}) corresponds to the
$N$ instances of $\psi$
in (\ref{optclone}),
while the $(a^\dagger_1 b^\dagger_1+a^\dagger_2 b^\dagger_2)^{M-N}$
corresponds to the $M-N$ instances of $S$.
The projection
onto the completely symmetric subspace is built in automatically
in (\ref{phcloner}) through the commutation properties of the bosonic
operators. This intuitive explanation can be checked by explicitly
evaluating the completely symmetrized component of (\ref{optclone}), and (\ref{phcloner}).
One verifies that the density matrices of
the $a$--modes are indeed identical in both cases. Thus we have
finally understood the formal equivalence between our stimulated
emission cloning procedures and the classical optimal cloners.

\section{Conclusions and Outlook}

One may feel that with the remarks in the previous section the
work is really completed. Cloning via stimulated emission has been
shown to be realizable, and the formal reasons for its optimality
are understood. Furthermore, we proposed a concrete experimental
realization which should lead to results in the near future. The
initial intuition probably shared by many physicists that a gain
medium is something like a cloner was thus shown to be entirely
correct. Our study demonstrates the intimate connection between
the apparently deeply quantum-field-theoretical concept of
stimulated emission and the quantum-information concept of
cloning. It is the author's hope that there may be more things to
learn about quantum field theory by looking at it from a quantum-
information point of view.

Another moral of the present work refers to the technology of
quantum information. It reminds us that at least for specific
tasks there may sometimes be more natural and therefore possibly
also more practical implementations than quantum computing
networks. Concerning possible practical applications of the
present work it is worth mentioning that the optimal universal
cloner constitutes the optimal eavesdropping method in some
protocols for quantum cryptography \cite{bechmann}, so conceivably a
future Eve
could rely on stimulated emission. We have discussed a possible
implementation using parametric down-conversion in some detail.
Other implementations might be possible, most notably based on
combining cavity QED and Bose-Einstein condensation. It should be
mentioned that parametric down-conversion-like Hamiltonians can be
realized for BECs, which makes them natural candidates for the
implementation of the cloning of atomic states.

\chapter{The No--Signaling Condition}

\section{Introduction}
\label{sec1}

The special theory of relativity
is one of the cornerstones of our
present scientific world-view. One of its most important features
is the fact that there is a maximum velocity for signals, i.e. for
anything that carries information, identical to the velocity of
light in vacuum. Within the special theory of relativity,
superluminal communication would immediately lead to all kinds of
causal paradoxes, e.g. one would be able to influence one's own past.

Another cornerstone of our present understanding of the world is
quantum physics. Quantum physics seems to have ``nonlocal''
characteristics
due to quantum  entanglement. Most importantly, it is not compatible
with local hidden variables, as shown by the violation of
Bell's inequalities \cite{bell},
which has been experimentally confirmed in several experiments
\cite{aspect,weihs}.

It is very remarkable that in spite of its non-local features, quantum mechanics
is compatible with the special theory of relativity, if it is assumed that
operators referring to space-like separated regions commute. In particular, one cannot
exploit quantum-mechanical entanglement between two space-like separated
parties for communication of classical messages faster than light
\cite{ghirardi}.

This peaceful coexistence between quantum physics and special relativity has led physicists to ask
whether the
principle of the impossibility of superluminal communication, which we will refer to as the ``no-signaling
condition'', could be used as an axiom in deriving basic features of quantum
mechanics. Here we show that it is indeed possible. If the usual {\it kinematical}
characteristics of quantum mechanics are assumed, then its {\it dynamical} rules can be derived from the
no-signaling assumption. By quantum kinematics we mean the following: the states of our systems are
described by vectors in a Hilbert space, and the usual rules for the results of measurements apply,
including the projection postulate. However, no a priori assumption is made about the time evolution of
the system.

 Our result is then, more precisely, that under the stated conditions the dynamics of our
system has to be described by {\it completely positive (CP) linear} \cite{preskill} maps.
This is equivalent to saying that under the given assumptions quantum mechanics is essentially
the only option since according to the Kraus representation theorem \cite{preskill}, every
CP map can be realized by a quantum-mechanical process, i.e. by a unitary (linear) evolution
on a larger Hilbert space (while on the other hand any quantum process corresponds to a CP map).
This rather surprising result is an extension of earlier work by N. Gisin \cite{gisinhpa}.

In the following, we will first recall how superluminal communication is impossible in
quantum mechanics in spite of the existence of entangled states, as a consequence of the
linearity of quantum dynamics.
Then we show that quantum kinematics
and no-signaling imply quantum dynamics. The argument proceeds
in two steps. Firstly, it is shown that the existence of entanglement, the projection postulate
and the no-signaling constraint imply linearity. Secondly,
complete positivity
of the dynamics follows
from the the existence of entanglement and linearity.

\section{No--signaling in Quantum Mechanics}

Consider two parties, denoted by Alice and Bob, who are space-like separated,
which implies that all operations performed by Alice commute with all
operations performed by Bob. (Throughout this work we will assume that in {\it
this} sense locality is implemented in the quantum kinematics.) Can they use a
shared entangled state $\psi_{AB}$ in order to communicate in spite of their
space-like separation?

The short answer is: no, because the situation on Bob's side will always be described
by the same reduced density matrix, whatever Alice chooses to do.
All the effects of her operations (described by linear maps) disappear when her
system is traced over. This answer is correct, but not very detailed, and thus it may
not be entirely convincing. In particular, a question that is frequently raised in this context
is the following: Alice could choose to measure her system in two different bases and thus
project Bob's system into different pure states depending on the basis she chose and her
measurement result. Since it is possible to distinguish two different states in quantum mechanics,
at least with some probability, shouldn't it be possible for Bob to infer her choice of basis,
at least in some percentage of the cases (which would be dramatic enough)?

Of course, the answer is no again, for the following reason. In order to gain information
about which basis Alice chose to measure, Bob can only perform some (generalized) measurement
on his system. Then he has to compare the conditional probabilities for this result to occur,
for the case that Alice measured in the first or in the second basis. But these conditional
probabilities will always be exactly the same for both possibilities. This can be seen as
a consequence of the linearity of quantum mechanics: Suppose that Alice's first choice
projects Bob's system into states $\psi_i$ with probabilities $p_i$ and her second choice
projects it  into states $\phi_\mu$ with probabilities $q_\mu$. Bob can calculate
the probability for his obtained result in every one of the states, and then weight these
probabilities with the probability to have this specific state. But because of the linearity of
any operation that Bob can perform on his states during his generalized measurement
procedure, his final result will only depend on the density matrix of the probabilistic
mixtures, which is the same in both cases, because they were generated from the same entangled
state.
For an example how two such mixtures can become distinguishable through a non-linear
(non-quantum-mechanical) evolution, see \cite{gisinweinberg}.

Let us note that
this argument also implies the non-existence of a perfect cloner in quantum mechanics
because such a machine would allow superluminal communication \cite{herbert}.
The impossibility of perfect cloning can also be shown directly from the linearity of
quantum mechanics \cite{wzurek}.

\section{No-signaling and Linearity}
\label{sec3}
In this section we show how quantum dynamics can be derived
from quantum kinematics and the no-signaling condition.
By quantum kinematics we mean that
the usual Hilbert space--structure (including entanglement)
and the projection postulate are assumed. The probabilities
of the results of measurements are assumed to be determined
by the density matrices of the systems in the usual way.
Thus if we consider a subsystem of the whole Universe
it will in general be in an entangled state with other parts
of the Universe. In particular, it may also happen that a system
denoted by $A$ is entangled with another system $B$ which
is {\em space-like} separated with respect to $A$, such that
their observable algebras do commute.
This is where the no-signaling constraint comes into play.
The dynamics of the systems has to be such that in spite of this entanglement no superluminal
communication between $A$ and $B$ is possible.

Suppose that $A$ and $B$ together are in the entangled state
$|\psi\rangle_{AB}$ with reduced density matrix $\rho_A$ for system
$A$.
As a consequence of the projection postulate,
by performing a measurement of his system  the observer $B$ also prepares a
certain state in $A$. In particular, {\em every} probabilistic mixture of
pure states corresponding to the density matrix $\rho_A$ can
be prepared via appropriate measurements on $B$
(for a proof see Sec. (\ref{ghjw}) and Ref.~\cite{hughston}).

Consider two such probabilistic mixtures
$\{P_{\psi_i}, x_i\}$
and $\{P_{\phi_j}, y_j\}$, where
$P_{\psi_k}$ is the projector corresponding to the pure state
$|\psi_k\rangle$ and $x_k$ is its probability,
such that
\be
\sum \limits_i x_i
P_{\psi_i}=\sum \limits_j y_j P_{\phi_j}=\rho_A.
\label{xx}
\ee
According to the no-signaling principle there should be no
way for the  observer in $A$ to distinguish these different probabilistic
mixtures.

A general dynamical evolution in system $A$ is of the form
\be
g: P_{\psi}\rightarrow g(P_{\psi})
\ee
where, most importantly, $g$ is not necessarily linear. Furthermore,
$g(P_{\psi})$ does not have to be
a pure state. Firstly, it could evolve into a mixed state. Secondly, if $\psi$ evolves into a
probabilistic mixture, then we define $g(P_{\psi})$ to denote the
corresponding density matrix.
Under such dynamics the probabilistic mixture
$\{P_{\psi_k}, x_k\}$  goes into another probabilistic mixture
$\{g(P_{\psi_k}), x_k\}$. Therefore the two final density
matrices after the action of $g$ on two different probabilistic mixtures
$\{P_{\psi_i}, x_i\}$ and $\{P_{\phi_j}, y_j\}$
are
\begin{eqnarray}
\rho_A^\prime \{P_{\psi_i}, x_i\} = \sum \limits_i x_i
g(P_{\psi_i})
\nonumber\\
\rho_A^\prime \{P_{\phi_j}, y_j\} = \sum \limits_j y_j
g(P_{\phi_j})
\end{eqnarray}
which {\em a priori} can be different. Let us recall that according
to our assumptions the results of all measurements in $A$ are determined
by the reduced density matrix $\rho_A^\prime$. This means
that as a consequence of the no-signaling principle
the density matrix $\rho_A^\prime$ at any later time has to be the
same for all probabilistic mixtures corresponding to a given
initial density matrix $\rho_A$. That is, it has to be a function
of $\rho_A$ only.

We can therefore write
\be
\rho_A^\prime =
g(\rho_A)=g(\sum \limits_i x_i P_{\psi_i}).
\ee
From the above it also follows that
\be
\rho_A^\prime =
\sum \limits_i x_i g(P_{\psi_i}),
\ee
therefore $g$ satisfies the condition
\be
g(\sum \limits_i x_i P_{\psi_i})=\sum \limits_i x_i
g(P_{\psi_i}),
\label{gofrho}
\ee
which implies that the map $g$ is {\em linear}.
Let us stress that there are three crucial ingredients
in our argument:
the existence of entanglement, the projection postulate,
and the no-signaling condition.
Specifically, the projection postulate
leads to  probabilistic mixtures and thus to the right-hand side
of Eq. (\ref{gofrho}). On the other hand, the no-signaling
condition tells us that the dynamics can depend only on the reduced
density matrix, which leads to the left-hand side of
Eq. (\ref{gofrho}).
Positivity is necessary in order to ensure that $g(\rho_A)$ is again
a valid density matrix, i.e. to ensure the positivity of all probabilities
calculated from it.

As we have made no specific assumptions about the system $A$
(apart from the fact that it can be entangled with some other system),
this means that the dynamics of our theory has to be linear
in general.

Let us now argue that linearity and positivity
already imply {\em complete positivity} in the present
framework.
To see this, consider again two arbitrary subsystems $A$ and $B$
which may again be in an entangled state $|\psi\rangle_{AB}$.
Now it is conceivable that system $A$ is changed locally
(i.e. the system evolves, is measured etc.), which is
described by some operation $g_A$, while {\em nothing} happens in $B$.
This formally corresponds to the operation
$g_A\otimes\openone_B$ on the whole system. Strictly speaking, we have made an additional assumption here,
namely that the identity operation
$\openone_B$ on a subsystem is a valid physical operation.

The joint operation $g_A\otimes\openone_B$ should take the
density matrix of the composite system $\rho_{AB}$ into
another valid (i.e. positive) density matrix, whatever the dimension
of the system $B$. But this is exactly the definition
of {\em complete positivity} for the map $g_A$ \cite{preskill}. If $g_A$ is
positive but not CP, then by definition there is always
some entangled state $\rho_{AB}$ for which
$g_A\otimes\openone_B (\rho_{AB})$ is no longer a positive
density matrix and thus leads to unphysical results such as
negative probabilities. Let us recall that transposition of system $A$
is an example for a positive but non-CP map.

In this way the existence of entangled states and the requirements
of positivity and linearity actually force us to admit
only completely positive dynamics.
As mentioned already in the introduction, this is equivalent to saying that under the given
assumptions quantum dynamics is essentially the {\em only} option since any
CP map can be realized by a quantum mechanical process, and on the other
hand, any quantum-mechanical process corresponds to a CP map.

Let us recall once again our starting assumptions: these were the existence of
entanglement, the projection postulate, the no-signaling
condition, and, strictly speaking, the assumption that the identity operation on
a subsystem is a permitted dynamical evolution. Nonlinear modifications of
quantum mechanics \cite{Czachor97}
have to give up at least one of these assumptions.
For instance, if the dynamics is allowed to depend only on the
reduced density matrix $\rho_A$, but in a nonlinear way,
then it is clear that $\rho_A$ cannot correspond to a probabilistic
mixture of pure states. But $\rho_A$ will correspond to such a
mixture whenever the observer in $B$ chooses to make an appropriate
measurement, as long as we believe in the projection postulate.
This implies that the projection postulate has to be modified
in such a nonlinear theory.
Another example would be a theory where some entangled states
are {\em a priori} excluded from the kinematics. In this case
some non-CP maps might be permissible. An extreme example would
be a theory without entanglement. Such a theory would of course be
in conflict with experiments.
An example for a linear, positive, but non-CP map consistent with
the no-signaling condition is the transposition of the density matrix
of the whole Universe. However in this case the identity operation
on a subsystem is not an allowed dynamics.

\section{Preparation of any mixture at a distance}
\label{ghjw}

Let us now show that any mixture corresponding to a given density matrix
can be prepared at a distance from any entangled state with the correct
reduced density matrix \cite{gisinhpa,hughston}.
Let us denote the system under consideration
by $A$ and the remote system by $B$. An immediate requirement on the state
of the joint system $|\psi\rangle_{A B}$ in order to achieve this is that it
needs to have the correct reduced density matrix $\rho_A$. Let
us denote the eigenvector representation of $\rho_A$ by
\be
\sum_{k=1}^r \lambda_k|v_k\rangle\langle v_k|.
\ee
Then $|\psi\rangle_{A B}$ must have
a Schmidt decomposition
\be
|\psi\rangle_{A B}=\sum_{k=1}^r \sqrt{\lambda_k}\,|v_k\rangle |g_k\rangle,
\ee
where the $|g_k\rangle$ are orthonormal states of system $B$.
We want to show that any decomposition of $\rho_A$ as a mixture of
pure states can be prepared from this state by operations on system
$B$ only.
To this end, consider an arbitrary decomposition
\be
\rho_A=
\sum_{i=1}^m x_i|\psi_i\rangle\langle\psi_i|,
\ee
where in general $m>r$.
Clearly this decomposition could
be obtained from a state
\be
|\phi\rangle_{A B}=\sum_{i=1}^m\sqrt{x_i}\,|\psi_i\rangle|\alpha_i\rangle,
\ee
with the $|\alpha_i\rangle$ being an orthonormal basis of a $m$--dimensional
Hilbert space $H_m$.
It seems that we now require a larger Hilbert space
in location $B$ in order to accommodate all the orthonormal $|\alpha_i\rangle$.
But the state $|\phi\rangle_{A B}$ also has a Schmidt representation
\be
|\phi\rangle_{A B}=\sum_{k=1}^r\sqrt{\lambda_k}\,|v_k\rangle |h_k\rangle,
\ee
which implies that $|\phi\rangle_{A B}$ and $|\psi\rangle_{A B}$
are connected by a
unitary transformation on $B$ alone:
\be
 |\phi\rangle_{A B}=\openone_A\otimes U_B |\psi\rangle_{A B},
\ee
so the dimension of the support of the reduced density matrix $\rho_B$
is the same for both states.
This means that $|\phi\rangle_{AB}$ can be rewritten as
\be
|\phi\rangle_{A B}=P_S|\phi\rangle_{A B}=
\sum_{i=1}^m\sqrt{x_i}\,|\psi_i\rangle|\tilde{\alpha}_i\rangle,
\ee
where $|\tilde{\alpha}_i\rangle=P_S|\alpha_i\rangle$ and
$P_S$ is the projector onto the support of $\rho_B$, i.e. the span
of the vectors $|h_k\rangle$.
The vectors $|\tilde{\alpha}_i\rangle$ live in this $r$--dimensional
span. Let us recall that $|\phi\rangle_{A B}$ can be generated from
$|\psi\rangle_{A B}$
by the local unitary $U_B$.

The states $|\psi_i\rangle$ can now be prepared with the help of the
POVM
\be
\sum_{i=1}^m|\tilde{\alpha}_i\rangle\langle\tilde{\alpha}_i|=1
\ee
on system $B$. Note that the $|\tilde{\alpha}_i\rangle$
form a POVM by construction:
They can be extended  to the orthonormal basis $\{|\alpha_i\rangle\}$
on the larger Hilbert space $H_m$. The proof of our above statement uses
exactly this extension. For a general POVM $\sum P_l=1$ on system $B$
the result $i$ prepares the state $\rho_A^i=\mbox{Tr}_B (P_i)_B\rho_{AB}$
in system $A$. Therefore the state prepared by the result corresponding to
$|\tilde{\alpha}_i\rangle$  is equal to $\rho_{A}^i=\mbox{Tr}_B
|\tilde{\alpha}_i\rangle\langle\tilde{\alpha}_i|
|\phi\rangle_{AB}\langle\phi|_{AB}$,
which can be rewritten, formally extending the Hilbert space dimension
for system $B$, as
\be
\mbox{Tr}_B P_S|\alpha_i\rangle \langle\alpha_i|P_S|\phi\rangle_{AB}
\langle\phi|=
\mbox{Tr}_B |{\alpha}_i\rangle \langle{\alpha}_i||\phi\rangle_{AB}
\langle\phi|
\nonumber
\\
=x_i|\psi_i\rangle\langle\psi_i|.
\ee
This completes the proof of our above statements \cite{preskill}.
Any decomposition of $\rho_A$ in a mixture of pure states can be obtained
from the state $|\psi\rangle_{AB}$.

\section{Bounds on cloning from positivity and linearity}

The motivation for the present section is partially historical. In \cite{gisinbound}
the no-signaling condition in the form of \cite{gisinhpa}, which was just positivity and linearity,
was used to derive a bound on the simplest cloner. According to our above argumentation, {\it all} of
quantum dynamics can be derived from the no-signaling condition, so the bounds from no-signaling are
identical to the quantum-mechanical bounds, which in the case of cloning are known \cite{brussprl,werner}.
Nevertheless, the method of \cite{gisinbound} is quite convenient for deriving bounds on universal quantum
machines \cite{simbuzgis}. We illustrate it here for the case of $1\rightarrow N$ cloning, thus providing
an alternative (maybe more direct) derivation of the corresponding bounds. We will first recall Gisin's
treatment of the $1\rightarrow 2$ cloner, then we generalize to the $1\rightarrow N$ case.

Let the initial state of the input qubit be denoted by
$\rho_0=\frac{1}{2}(1+\vec \sigma \cdot \vec m)=|+\vec m\rangle
\langle +\vec m|$, where $\vec \sigma \cdot \vec m |+\vec
m\rangle=+|\vec m\rangle$. The output two-qubit density matrix is
denoted by $\rho(\vec m)$. From the discussion of the preceding
section we know that $\rho(\vec m)$ has to be a linear function of
$\vec m$.

The output density matrix $\rho(\vec m)$ is further constrained by
the requirement of universality, which takes the form
\begin{equation}
\rho(U\vec m)=U\otimes U\rho(\vec m)U^\dagger\otimes U^\dagger
\end{equation}
for all unitary operators U. This  implies that
$\rho(\vec m)$ depends only on $\vec m$ and on no other privileged
direction. Thus, if $\rho(\vec m)$ is written in the basis of
matrices
\begin{equation}
\openone\otimes\openone,\openone\otimes\sigma_i,\sigma_i\otimes\openone,\sigma_i\otimes\sigma_k,
\end{equation}
the coefficients can only depend on the components $m_i$ of $\vec
m$ and on the invariant tensors $\delta_{ij}$ and
$\epsilon_{ijk}$.

It is thus necessarily of the form: \beq \rho(\vec
m)=\frac{1}{4}\left(\openone\otimes\openone + \eta_1\vec
m\vec\sigma\otimes\openone + \eta_2\openone\otimes\vec m\vec\sigma
+ t~\vec\sigma\otimes\vec\sigma +t_{xy}\vec
m(\vec\sigma\wedge\vec\sigma) \right) \eeq where
$\eta_1,\eta_2,t,t_{xy}$ are real parameters. In order for
$\rho(\vec m)$ to be a physical density matrix, its eigenvalues
have to be non-negative. A simple calculation shows that this
implies \beqa 1+t\pm(\eta_1+\eta_2)&\ge&0 \nonumber
\\ 1-t\pm\sqrt{4t^2+4t_{xy}^2+(\eta_1-\eta_2)^2}&\ge&0 \label{pos}\eeqa

In the case of symmetric cloning, the task is to optimize the
fidelity $F=Tr(\rho(\vec m)~P_{\vec m}\otimes\openone)$, where
$P_{\vec m}=|+\vec m\rangle \langle +\vec m|$, assuming
$\eta_1=\eta_2\equiv\eta$.  A simple calculation leads to the
optimal values $t_{xy}=0,t=1/3,\eta=2/3$, for which
$F=\frac{5}{6}$ . Note that this also optimizes $Tr(\rho(\vec
m)~P_{\vec m}\otimes P_{\vec m})=\frac{2}{3}$. These are exactly
the bounds that are valid in quantum mechanics, cf. Sec. \ref{sandc}.

Now we show, that the above result can be generalized to
the case of $1\rightarrow N$ cloning.
Firstly, from any non-covariant and non-permutation invariant
cloning transformation that produces a number of copies that are
scaled versions of the input qubit with identical scaling factors,
by averaging over unitary transformations and permutations one can
get a covariant and permutation invariant transformation without
affecting the quality of the copies \cite{keylwerner}.
This means that in deriving
bounds we can restrict our attention to covariant and permutation
invariant output density matrices.

Secondly, as shown above, it follows from no--signaling that
$\rho_{out}$ has to be linear in the Bloch vector
of the input qubit.

Thus for building the output density
matrix we are only left with $m_i$ and $\delta_{jk}$ as possible
coefficients in the Pauli matrix representation, where $m_i$
can only occur linearly. The invariant tensor $\epsilon_{ijk}$ is
excluded by the requirement of
permutation invariance. Possible terms are:
\begin{eqnarray}
&\openone \otimes \openone \otimes... \otimes \openone& \nonumber\\
&\sigma_k \otimes \sigma_k \otimes \openone \otimes ... \otimes
\openone+ perm.&\nonumber\\&\sigma_k \otimes \sigma_k \otimes
\sigma_l \otimes \sigma_l \otimes \openone \otimes ... \otimes
\openone+ perm.&\nonumber\\ &....& \nonumber\\ &\vec{\sigma} \cdot
\vec{m} \otimes \openone \otimes ... \otimes \openone +
perm.&\nonumber\\ &\vec{\sigma} \cdot \vec{m} \otimes \sigma_k
\otimes \sigma_k \otimes \openone \otimes ... \otimes \openone +
perm.&\nonumber\\ &....&,
\end{eqnarray}
where summation over repeated indices is understood. Because of
universality we can choose $\vec{m}$ e.g. along the z-axis. Then
one can convince oneself that all the above terms can be generated
by products of
\begin{equation}
J_z=\frac{1}{2} (\sigma_z \otimes \openone \otimes ... \otimes
\openone + ...)
\end{equation}
and
\begin{equation}
\vec{J}^2=J_k J_k,
\end{equation}
the z-component of the total angular momentum and its square
respectively. $J_z$ can occur only linearly, while for $\vec{J}^2$
higher powers are possible: one has to distinguish the cases of
$N$ even and $N$ odd. For $N=2k$ $(\vec{J}^2)^n$ can go up to
$n=k$ and $J_z (\vec{J}^2)^n$ up to $n=k-1$, while for $N=2k+1$
 $(\vec{J}^2)^n$ can go up to
$n=k$ and $J_z (\vec{J}^2)^n$ up to $n=k$ as well. Higher powers
are linearly dependent. We will denote the maximum possible values
as $n_{max}$ and $n_{max}'$ in the following. Note that
$n_{max}+n_{max}'=N-1$.

Therefore the most general output density matrix can be written in
the following way:
\begin{equation}
\rho_{out}=\beta_{0} \openone + \sum \limits_{n=1}^{n_{max}}
\beta_n (\vec{J}^2)^n + \alpha_0 J_z + \sum
\limits_{n=1}^{n_{max}'} \alpha_n J_z (\vec{J}^2)^n,
\end{equation}
where $\openone$ now denotes the unit matrix in the $N$-particle
Hilbert space. Our task is to find coefficients $\alpha_i$ and
$\beta_i$ such that the scaling factor of an individual clone is
maximal. The constraints are given by the requirements of
positivity and normalization: all eigenvalues of $\rho_{out}$ have
to be positive, and its trace has to be equal to unity.

In order to express the positivity constraint one has to
diagonalize the matrix, but in the present formulation this is
trivial. The eigenvectors are just the angular momentum
eigenvectors $|j,m,\gamma_j\rangle$, where $\gamma_j$ runs over
the different irreducible representations for a given $j$, i.e.
$\gamma_j=1,...,d_j$, where $d_j$ denotes the number of irr. reps
for $j$. This means that the eigenvalues are given by
\begin{equation}
\lambda(j,m)=\beta_{0} + \sum \limits_{n=1}^{n_{max}} \beta_n
(j(j+1))^n + \alpha_0 m + \sum \limits_{n=1}^{n_{max}'} \alpha_n m
(j(j+1))^n. \label{eigenv}
\end{equation}
(The eigenvalues do not depend on $\gamma_j$.) Positivity implies
that
\begin{equation}
\lambda(j,m)\geq0 \hspace{1cm} \forall j,m.
\end{equation}
The normalization constraint is
\begin{equation}
\mbox{Tr} \rho_{out}=\beta_0 2^N + \sum \limits_{n=1}^{n_{max}}
\beta_n \mbox{Tr} (\vec{J}^2)^n=1,
\end{equation}
because the trace of the terms with $J_z$ is zero. This can be
expressed as
\begin{equation}
\beta_0 2^N + \sum \limits_{n=1}^{n_{max}} \sum
\limits_{j=j_{min}}^{N/2} \beta_n d_j (2j+1) (j(j+1))^n=1.
\label{norm}
\end{equation}
For $N=2k$ $j_{min}=0$, while for $N=2k+1$ $j_{min}=1/2$.

We still have to determine the scaling factor of the individual
clones, which is the quantity that we want to maximize. This
requires calculation of the one-particle reduced density matrix,
which in our case must have the form
\begin{equation}
\mbox{Tr}_{N-1} \rho_{out}=\frac{1}{2}(\openone+s \sigma_z).
\end{equation}
We want to maximize the coefficient of $\sigma_z$. The terms in
$\rho_{out}$ leading to a $\sigma_z$ are
\begin{equation}
\alpha_0 \mbox{Tr}_{N-1}J_z + \sum \limits_{n=1}^{n_{max}'}
\alpha_n \mbox{Tr}_{N-1} J_z (\vec{J}^2)^n.
\end{equation}
The scaling factor $s$ is obtained by multiplication with
$\sigma_z$ and tracing over the remaining particle. Using
\begin{equation}
\mbox{Tr}_1 \left( \sigma_z \mbox{Tr}_{N-1} J_z (\vec{J}^2)^n
\right) =\frac{2}{N} \mbox{Tr}_N (J_z)^2 (\vec{J}^2)^n,
\end{equation}
\begin{equation}
\mbox{Tr}_N (J_z)^2 (\vec{J}^2)^n=\sum \limits_{j=j_{min}}^{N/2}
d_j (j(j+1))^n \sum \limits_{m=-j}^{j} m^2,
\end{equation}
and
\begin{equation}
\sum \limits_{m=-j}^j m^2=\frac{1}{3}j(j+1)(2j+1)
\end{equation}
 one obtains
\begin{equation}
s=\alpha_0 2^{N-1} + \frac{2}{3N} \sum \limits_{n=1}^{n_{max}'}
\alpha_n \sum \limits_{j=j_{min}}^{N/2} d_j j(j+1)(2j+1)
(j(j+1))^n. \label{scalf}
\end{equation}
By the structure of Eqs. (\ref{eigenv}),(\ref{norm}), and
(\ref{scalf}) one is led to make the substitution
\begin{eqnarray}
a_j=\sum \limits_{n=1}^{n_{max}'} \alpha_n (j(j+1))^n \hspace{1cm}
j=j_{min},...,N/2\nonumber\\ b_j=\sum \limits_{n=1}^{n_{max}}
\beta_n (j(j+1))^n \hspace{1cm} j=j_{min},...,N/2.
\end{eqnarray}
Note that $a_0$ and $b_0$, which arise for even $N$ because
$j_{min}=0$, are identically zero. Note also that a priori this
does not seem to be a good change of variables for the
optimization because in general the $a_j$ and $b_j$ are not all
linearly independent, as one can see by counting their number and
comparing to the number of $\alpha_n$ and $\beta_n$. We will
discuss this problem in detail below when we present the real
change of variables made, the above substitution is only an
intermediate step.

In these variables the optimization problem has the following
form:
\begin{eqnarray}
&\beta_0 + b_j + (\alpha_0 + a_j) m \geq 0 \hspace{0,5cm} \forall
j,m&\nonumber\\ \nonumber\\&\beta_0 2^N + \sum
\limits_{j=j_{min}}^{N/2} b_j d_j (2j+1)=1&\\ &\alpha_0 2^{N-1}+
\frac{2}{3N} \sum \limits_{j=j_{min}}^{N/2} a_j d_j
j(j+1)(2j+1)=\mbox{Max.!}&\nonumber
\end{eqnarray}
Noting that
\begin{equation}
\sum \limits_{j=j_{min}}^{N/2} d_j (2j+1)=2^N,
\end{equation}
because it is the number of dimensions of all irreducible
representations, and that
\begin{equation}
\sum \limits_{j=j_{min}}^{N/2} d_j (2j+1)j(j+1) =
\mbox{Tr}(\vec{J}^2)=3N 2^{N-2},
\end{equation}
as can be checked by direct calculation of the trace, one is led
to make the further redefinition
\begin{eqnarray}
A_j=(\alpha_0+a_j)d_j(2j+1) \hspace{1cm} j=j_{min},...,N/2
\nonumber\\ B_j=(\beta_0+b_j)d_j(2j+1) \hspace{1cm}
j=j_{min},...,N/2.
\end{eqnarray}
In terms of the independent variables $\alpha_n$ and $\beta_n$
this reads
\begin{eqnarray}
A_j=d_j(2j+1)\left( \alpha_0+\sum \limits_{n=1}^{n_{max}'}
\alpha_n (j(j+1))^n \right)\hspace{1cm}
j=j_{min},...,N/2\nonumber\\
 B_j=d_j(2j+1)\left( \beta_0+\sum
\limits_{n=1}^{n_{max}} \beta_n (j(j+1))^n \right)\hspace{1cm}
j=j_{min},...,N/2  \label{changev}
\end{eqnarray}
Now we have to face the question whether this is a legal change of
variables, i.e. whether the $A_j$ and $B_j$ are linearly
independent. Let's first discuss the second line of
(\ref{changev}). There are $n_{max}+1$ independent parameters on
the right hand side, which is $k+1$ for $N=2k$ and also $k+1$ for
$N=2k+1$ (see above). This is identical to the number of different
possible values of $j$. This means that the number of $B_j$ is the
same as the number of $\beta_n$, the only question left is whether
the matrix connecting the two sets of variables is invertible.
This last point is easy to show. (It's determinant is a Van der
Monde determinant.)

Turning to the first line of (\ref{changev}) we see that the
number of independent parameters on the right hand side is
$n_{max}'+1$, which is $k$ for $N=2k$ and $k+1$ for $N=2k+1$. This
means that there seems to be a problem for the case $N=2k$,
because one of the $A_j$ is a linear combination of the others.
Fortunately it turns out, as we will see below, that the variable
$A_0$ does not play any role in the optimization, which allows us
to disregard it. The other $A_j$ for $j\neq0$ can be shown to be
linearly independent exactly as the $B_j$.

Having justified our change of variables, we can now study its
consequences. It leads to the following set of conditions:
\begin{eqnarray}
&\lambda(j,m)=B_j+A_j m \geq 0 \hspace{1cm} \forall j,m&
\nonumber\\ &\sum \limits_{j=j_{min}}^{N/2} B_j=1& \\
&\frac{2}{3N} \sum \limits_{j=j_{min}}^{N/2} A_j
j(j+1)=\mbox{Max.!}& \nonumber \label{optimize}
\end{eqnarray}
From the first and third line one sees that $A_0$ only enters
multiplied by zero and therefore doesn't play any role, as
mentioned above. From the first line it follows that the $B_j$
have to be positive, and also that $j|A_j|\leq B_j$. From the
third line it is clear that negative values of $A_j$ are not
helpful, therefore one obtains
\begin{equation}
jA_j \leq B_j \hspace{0,5cm}\forall j.
\end{equation}
If all these inequalities are saturated, one gets
\begin{equation}
s=\frac{2}{3N} \sum \limits_{j=j_{min}}^{N/2} B_j (j+1).
\end{equation}
From the above it is clear that the maximum is obtained if the
value of $B_j$ for the largest possible $j$, i.e. for $j=N/2$, is
equal to unity, with all other $B_j$ equal to zero. This leads to
\begin{equation}
s_{max}=\frac{1}{3}+\frac{2}{3N},
\end{equation}
which is exactly the maximum possible scaling factor in quantum
mechanics.

\section{Conclusions}

We find it quite remarkable that what we referred to as quantum
kinematics, i.e. the Hilbert space structure and the projection postulate,
together with the no-signaling condition already constrains the dynamics
to be of the form that we know from quantum mechanics: linear and
completely positive. Concerning the challenge to truly derive quantum
mechanics from some fundamental principles, the present result is
certainly just a small piece of the puzzle. However, besides providing
some insight into the interrelations between different properties of the
standard theory, this result also leads to a clear statement about
possible non-linear modifications of quantum mechanics, namely that they
have to give up at least one of the assumptions made in our derivation.
Although the author has some sympathy for the program of studying
non-linear extensions, at present he is not
sure which assumption he would be most willing to give up.

\chapter{A Simple Kochen-Specker Experiment}

\section{Introduction}

Most predictions of quantum mechanics are of a statistical nature,
with the theory making probabilistic predictions for individual
events. The question whether one can go beyond quantum mechanics
in this respect, i.e. whether there could be hidden variables
determining the results of all individual measurements, has been
answered to the negative for {\it local} hidden variables by
Bell's theorem \cite{bell}. Locality means that in such theories
the results of measurements in a certain space-time region are
independent of what happens in a space-time region that is
space-like separated, in particular independent of the settings of
a distant measuring apparatus.

Bell's theorem refers to a situation where there are two particles
and where the predictions of quantum mechanics are statistical.
Furthermore, even definite (non-statistical) predictions of
quantum mechanics are in conflict with a local realistic picture
for systems of three particles or more \cite{ghz,ghzajp}.

The Kochen--Specker (KS) theorem \cite{specker,ks,bellks,peresbook}
states that {\it
non--contextual} hidden variable theories are incompatible with
quantum mechanics. Non-contextuality (NC) means that the value for
an observable predicted by such a theory does not depend on the
experimental context, i.e. which other co-measurable observables
are measured simultaneously.

To put the Kochen-Specker (KS) theorem in a proper context, let us
briefly recall some basic facts about measurements in quantum
mechanics. Let us first discuss sequential measurements. Consider
the sequential measurement of two commuting observables $A$ and $B$.
Ideally the same values for $A$ and $B$ are found again and again for
repeated measurements, provided that they are projective. A
measurement of $B$ does not seem to disturb the value of $A$. The
quantum mechanical explanation for this phenomenon is that the
first measurement of $A$ and the first measurement of $B$ project the
system into a joint eigenstate or eigenspace of $A$ and $B$ in which
it remains.

For sequential measurements of non-commuting observables the
situation is radically different. For example, consider repeated
measurements of $\sigma_z$ and $\sigma_x$. When measuring the sequence
$\sigma_z$, $\sigma_x$, $\sigma_z$ one may find $\sigma_z$ equal to $-1$ in
the third measurement after having found $\sigma_z$ equal to $+1$
in the first one. This can be visualized with polarizers and
photons. This shows that if there are hidden values for quantum
mechanical observables they must necessarily be affected by the
measurement. But originally, before a measurement was performed,
there might still have been one precise value for every
observable, which was then influenced by the first measurement
performed on the system.

Let us now consider joint measurements of several observables. In
quantum mechanics only joint measurements of commuting observables
are meaningful, because only these have joint eigenstates onto
which the measurement can project. One can ask the following
question: Is it possible that for all observables there are hidden
values that do no depend on which other observables are measured
jointly? For sequential measurements we have already seen that
measurements of non-commuting observables have an effect. But here
the situation is more subtle. We are talking about a situation
where some observable $A$ could be measured jointly with $B$ or with $C$
and we ask whether there can be an underlying theory such that the
value for $A$ does not depend on whether $B$ or $C$ are measured
jointly, and such that this holds for all $A$, $B$ and $C$. The KS
theorem states that there can be no hidden values of this kind:
measurements of commuting observables also matter.

The KS theorem was an important station on the road leading to
Bell's theorem. While one can argue that there is no very good
justification for expecting non-contextuality, if $A$, $B$ and $C$ are
all measured on a single particle, as in the original formulation
of the KS theorem, this changes dramatically, if $A$ on the one hand
and $B$ and $C$ on the other hand can be measured on two particles in
entirely different locations. One can say that Bell's discovery
was that the KS result remains true in such a situation as well.
For the hypothetical hidden values measurements of commuting
observables matter, even if they commute because of space-like
separation.

Let us briefly recall the setting of the original KS theorem. KS
considered a single spin-one particle, the relevant observables
are the squares of the spin components along arbitrary directions,
denoted e.g. by $S_x^2$ for the direction $x$. These observables
commute for orthogonal directions. They satisfy the constraint
\be S_x^2+S_y^2+S_z^2=s (s+1)= 2\ee\label{sumistwo} for all orthogonal triplets of directions
$\{x,y,z\}$.

The question of non-contextuality now poses itself in the
following way: is it possible to assign values 0 and 1 to all
directions such that the constraint (\ref{sumistwo}) is fulfilled? Thus the
question of the existence of non-contextual hidden values becomes
a coloring problem on the sphere. The non-existence of such a
coloring can be inferred from Gleason's theorem \cite{gleason}. Kochen
and Specker gave a direct proof by exhibiting a finite set of
directions (originally 117) that cannot be colored. Since then
proofs that require only smaller numbers of directions have been
found, see e.g. Ref. \cite{peresks}.

It is well known that Bell's theorem leads to possible
experimental tests of local hidden variables by studying the
violation of certain inequalities for correlation functions. While
tests of local hidden variables can also be seen as tests of
non-contextuality, as briefly explained above, so far there has
not been an experiment based on the original form of the
Kochen-Specker theorem. This would require testing that the
constraint Eq. (\ref{sumistwo}) is indeed fulfilled
for all directions belonging to the
Kochen-Specker set. But the message of the KS theorem is
weaker than that of Bell's theorem: non-contextual hidden
variables are a smaller class than local hidden variables. It
should therefore be possible to find an experiment disproving
non-contextuality that is considerably simpler than the usual
tests of Bell's inequalities. Furthermore, from the theoretical
point of view, the KS argument is quite elaborate. It should be
possible to reach the same conclusions in a much simpler way. Note
that the GHZ argument can already be seen as a much
simplified KS theorem. In the present chapter we show that the
above programme can be realized. We present a simple argument
against non-contextual theories which involves just a few
observables and leads to a simple experiment. The present work was
inspired by the work of Cabello and Garc\'{\i}a-Alcaine (CG)
\cite{cab}.

The experiment can be realized with single particles, using their
path and spin degrees of freedom. It leads to a non-statistical
test of non-contextuality versus quantum mechanics. In this
respect it is similar to the GHZ argument against local realism.

If the experiment is realized with photons, the setup that we
shall present only requires a source for single photons (such as
parametric down-conversion) and passive optical elements. In the
following, we first show how a very direct experimental test of
non-contextuality can be found, then we discuss our operational
realization.

\section{A Simple Kochen--Specker Argument \dots}

Consider four binary observables $Z_1, X_1, Z_2$, and $X_2$. Let
us denote the two possible results for each observable by $\pm 1$.
In a non-contextual hidden variable (NCHV) theory these
observables have predetermined non-contextual values $+1$ or $-1$
for individual systems, denoted as $v(Z_1), v(Z_2), v(X_1)$, and
$v(X_2)$. This means e.g. that for an individual system the result
of a measurement of $Z_1$ will always be $v(Z_1)$ irrespective of
which other compatible observables are measured simultaneously.

Now imagine an ensemble E of systems for which one always finds
equal results for $Z_1$ and $Z_2$, and also for $X_1$ and $X_2$.
(Clearly, in order for this statement to be meaningful, $Z_1$ and
$Z_2$, and $X_1$ and $X_2$ have to be co-measurable.) In the NCHV
theory this means that
\begin{equation}
 v(Z_1)=v(Z_2)\hspace{5mm}\mbox{and}\hspace{5mm} v(X_1)=v(X_2)
 \label{ensemble}
\end{equation}
 for each individual system of the
ensemble. Then there are only two possibilities: either
$v(Z_1)=v(X_2)$, which implies $v(X_1)=v(Z_2)$; or $v(Z_1)\neq
v(X_2)$, which implies $v(X_1) \neq v(Z_2)$. We will see that this
elementary logical deduction is already sufficient to establish a
contradiction between NCHV theories and quantum mechanics.

To this end, let us express the above argument in a slightly
different way. Eq. (\ref{ensemble}) can be written as
\begin{equation}
v(Z_1)v(Z_2)=v(X_1)v(X_2)=1.
\end{equation}
Multiplying by $v(X_2)v(Z_2)$ it immediately follows that
\begin{equation}
v(Z_1)v(X_2)=v(X_1)v(Z_2). \label{pred1}
\end{equation}
Let us now introduce the notion of product observables such as
$Z_1X_2$. By definition, one way of measuring $Z_1X_2$ is to
measure $Z_1$ and $X_2$ separately and multiply the results; in
general, there are other ways. In particular, if another
compatible observable (e.g. $X_1Z_2$, cf. below) is measured
simultaneously, it will in general not be possible to obtain
separate values for $Z_1$ and $X_2$. However, in a non-contextual
theory, the result of a measurement of an observable must not
depend on which other observables are measured simultaneously.
Therefore the predetermined value $v(Z_1X_2)$, for example, in a
NCHV theory has to follow the rule \cite{cab}
\begin{equation}
v(Z_1X_2)=v(Z_1)v(X_2).
\end{equation}
In this new language, our above argumentation can be resumed in
the following way:
\begin{equation}
 v(Z_1Z_2)=v(X_1X_2)=1\Rightarrow v(Z_1X_2)=v(X_1Z_2)
\label{pred}
\end{equation}
i.e. if our systems have the property expressed in Eq.
(\ref{ensemble}), then the two product observables $Z_1X_2$ and
$X_1Z_2$ must always be equal in a NCHV theory. Note that in
general this prediction of NCHV can only be tested if $Z_1X_2$ and
$X_1Z_2$ are co-measurable.

It follows from the results of \cite{cab} that the prediction
(\ref{pred}) leads to an observable contradiction with quantum
mechanics. To see this, consider a system of two qubits and the
observables \cite{cab}
\begin{equation}
Z_1:=\sigma_z^{(1)}, X_1:=\sigma_x^{(1)},
 Z_2:=\sigma_z^{(2)}, X_2:=\sigma_x^{(2)},
\end{equation}
  where
$\sigma_z^{(1)}$ means the z-component of the ``spin'' of the
first qubit etc. It is easy to check that this set of observables
satisfies all the properties required above. In particular, while
$Z_1$ and $X_1$, and $Z_2$ and $X_2$, do not commute, the two
product observables $Z_1X_2$ and $X_1Z_2$ do. Furthermore, the
quantum-mechanical two-qubit state
\begin{eqnarray}
|\psi_1\rangle&=&\frac{1}{\sqrt{2}}(|+z\rangle|+z\rangle+|-z\rangle|-z\rangle)\nonumber\\
&=&\frac{1}{\sqrt{2}}(|+x\rangle|+x\rangle+|-x\rangle|-x\rangle)
\label{state}
\end{eqnarray}
 is a joint eigenstate of the
commuting product observables $Z_1Z_2$ and $X_1X_2$ with both
eigenvalues equal to $+1$. Therefore, on the one hand the ensemble
described by this state possesses the property of the ensemble $E$
discussed above (cf. (\ref{ensemble})): the measured values of
$Z_1Z_2$ and $X_1X_2$ are equal to $+1$ for every individual
system. On the other hand, quantum mechanics predicts for the
state $|\psi_1\rangle$, that the measured value of $Z_1X_2$ will
always be opposite to the value of $X_1Z_2$. This can be seen by
decomposing $|\psi_1\rangle$ in the basis of the joint eigenstates
of the two commuting product observables $Z_1X_2$ and $X_1Z_2$:
\begin{equation}
|\psi_1\rangle=\frac{1}{\sqrt{2}}(|\chi_{1,-1}\rangle+|\chi_{-1,1}\rangle),
\label{sup}
\end{equation}
with

\begin{eqnarray}
|\chi_{1,-1}\rangle&=&\frac{1}{2}(|+z\rangle|+z\rangle+|-z\rangle|-z\rangle\nonumber\\
&&+|+z\rangle|-z\rangle-|-z\rangle|+z\rangle)\nonumber\\
&=&\frac{1}{\sqrt{2}}(|+z\rangle|+x\rangle-|-z\rangle|-x\rangle)\nonumber\\
&=&\frac{1}{\sqrt{2}}(|-x\rangle|+z\rangle+|+x\rangle|-z\rangle)\nonumber
\end{eqnarray}
\begin{eqnarray}
|\chi_{-1,1}\rangle&=&\frac{1}{2}(|+z\rangle|+z\rangle+|-z\rangle|-z\rangle\nonumber\\
&&-|+z\rangle|-z\rangle+|-z\rangle|+z\rangle)\nonumber\\
&=&\frac{1}{\sqrt{2}}(|+z\rangle|-x\rangle+|-z\rangle|+x\rangle)\nonumber\\
&=&\frac{1}{\sqrt{2}}(|+x\rangle|+z\rangle-|-x\rangle|-z\rangle).
\end{eqnarray}

and
\begin{eqnarray}
Z_1X_2 |\chi_{1,-1}\rangle = +|\chi_{1,-1}\rangle, X_1Z_2
|\chi_{1,-1}\rangle
=
-|\chi_{1,-1}\rangle\nonumber\\ Z_1X_2 |\chi_{-1,1}\rangle =
-|\chi_{-1,1}\rangle, X_1Z_2 |\chi_{-1,1}\rangle =
+|\chi_{-1,1}\rangle \label{eigenv_k}
\end{eqnarray}
From (\ref{sup}) and (\ref{eigenv_k}) one sees that $|\psi_1\rangle$
is a linear combination of exactly those joint eigenstates of
$Z_1X_2$ and $X_1Z_2$ for which the respective eigenvalues are
opposite, which means, of course, that in a joint measurement the
two observables will always be found to be different. With Eq.
(\ref{pred}) in mind, this implies that the ensemble described by
$|\psi_1\rangle$ cannot be described by any non-contextual hidden
variable theory.

Note that one would already have a contradiction if quantum
mechanics only predicted that the observed values of $Z_1X_2$ and
$X_1Z_2$ are sometimes different, but in fact the result is even
stronger, with QM and NCHV predicting exactly opposite results.

According to the argument presented in the previous paragraph, an
experimental test of non-contextuality can be performed in the
following way: (i) Show that $Z_1Z_2=1$ and $X_1X_2=1$ for systems
prepared in a certain way. (ii) Determine whether $Z_1X_2$ and
$X_1Z_2$ are equal for such systems. Note that in steps (i) and
(ii) the observables $Z_1, X_1, Z_2,$ and $X_2$ appear in two
different contexts.

Quantum mechanics predicts that step (i) can be accomplished by
constructing a source of systems described by the state
$|\psi_1\rangle$ and measuring $Z_1Z_2$ and $X_1X_2$ on these
systems. According to QM, both $Z_1Z_2$ and $X_1X_2$ will always
be found to be equal to $+1$. This can e.g. be verified by
measuring the pairs $Z_1$ and $Z_2$ and $X_1$ and $X_2$ separately
on many systems, and obtaining the values of $Z_1Z_2$ and $X_1X_2$
by multiplication. Alternatively, one could also perform joint
measurements of $Z_1Z_2$ and $X_1X_2$ on individual systems, but
for step (i) such joint measurements are not strictly necessary.
On the other hand, step (ii) definitely requires a joint
measurement of $Z_1X_2$ and $X_1Z_2$, because both negative and
positive values are to be expected for $Z_1X_2$ and $X_1Z_2$, and
we have to determine whether their values are equal or opposite
for individual systems.

\section{\dots Leading to a Possible Experiment}
One could consider realizing the above protocol with two
particles. However, since a joint measurement of the two qubits is
required it follows that locality is not an issue in the present
experiment. This suggests looking for a single-particle
realization for the sake of simplicity.

In our single-particle scheme, the first qubit is emulated by the
spatial modes of propagation (paths) of a single spin-1/2 particle
or photon, and the second qubit by its spin (or polarization)
degree of freedom \cite{marek,czachor}. Spin-1/2 and photon polarization
are completely equivalent for our purposes. Our setup requires a
source of polarized single particles, beam splitters, and
Stern-Gerlach type devices. In practice, the experiment would be
easiest to do with photons because all these elements are readily
available, in particular polarized single-photon states can be
produced to excellent approximation via parametric down-conversion
\cite{singlephot}.
Besides their conceptual simplicity, single-photon experiments are
attractive because very pure experimental conditions, in
particular very high visibilities, can be achieved.
Nevertheless, we will use the spin language in
the sequel because it is more familiar to most physicists.

Consider a situation where the particle can propagate in two
spatial modes $u$ and $d$, and let $|z+\rangle, |z-\rangle$ etc.
denote the particle's spin states as before. Then the state
$|\psi_1\rangle$ of Eq. (\ref{state}) is mapped onto the
one-particle state
\begin{equation}
\frac{1}{\sqrt{2}}(|u\rangle|z+\rangle+|d\rangle|z-\rangle).
\label{psi1}
\end{equation}
The observables $Z_1,X_1,Z_2,X_2$ are now represented by
\begin{eqnarray}
Z_1&=&|u\rangle\langle u|-|d\rangle \langle d|\nonumber\\
X_1&=&|u'\rangle\langle u'|-|d'\rangle \langle d'|\nonumber\\
Z_2&=&|z+\rangle\langle z+|-|z-\rangle \langle z-| \nonumber\\
X_2&=&|x+\rangle\langle x+|-|x-\rangle\langle x-|, \label{obs}
\end{eqnarray}
where $|u'\rangle=\frac{1}{\sqrt{2}}(|u\rangle+|d\rangle),
|d'\rangle=\frac{1}{\sqrt{2}}(|u\rangle-|d\rangle),|x+\rangle=
\frac{1}{\sqrt{2}}(|z+\rangle+|z-\rangle),
|x-\rangle=\frac{1}{\sqrt{2}}(|z+\rangle-|z-\rangle)$, i.e. $u'$
and $d'$ denote the output modes of a 50-50 beam-splitter with
inputs $u$ and $d$, and $|x+\rangle$ and $|x-\rangle$ are the spin
eigenstates along the $x$ direction . Clearly, $Z_1$ and $X_1$ act
on the path, and $Z_2$ and $X_2$ on the spin degree of freedom.

\begin{figure}
    \centering
        \includegraphics[width= 0.7 \columnwidth] {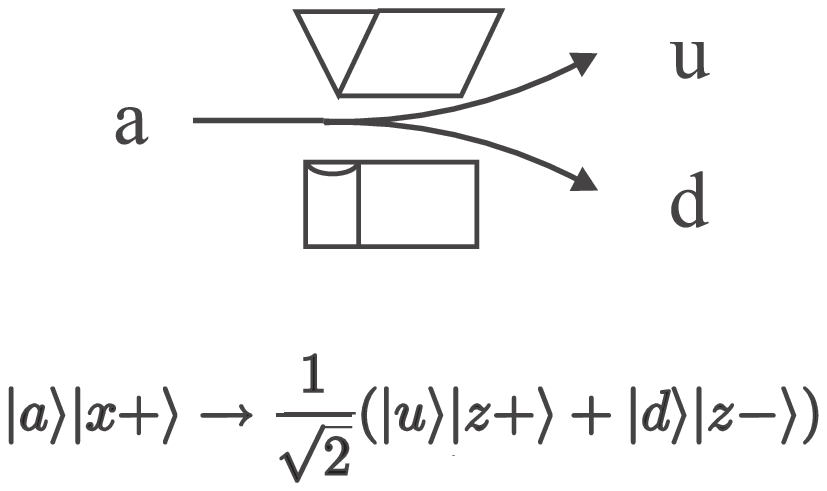}
        \caption{ Possible way of creating the single-particle version of $|\psi_1\rangle$
        given in Eq. (\ref{psi1}) using a standard Stern-Gerlach apparatus. A single particle with
        spin state $|x+\rangle=\frac{1}{\sqrt{2}}(|z+\rangle+|z-\rangle)$, i.e. spin
        along the positive $x$ direction, comes in from the left (spatial mode $|a\rangle$).
        By the Stern-Gerlach device, which separates incoming states according to the $z$-components
        of their spin, this is transformed into the desired superposition state.
        The outputs $u$ and $d$ could be
connected
        to the inputs of the devices of Figures \ref{separate} or \ref{joint}}
\label{source}
\end{figure}

\begin{figure}
    \centering
 \includegraphics[width=\columnwidth] {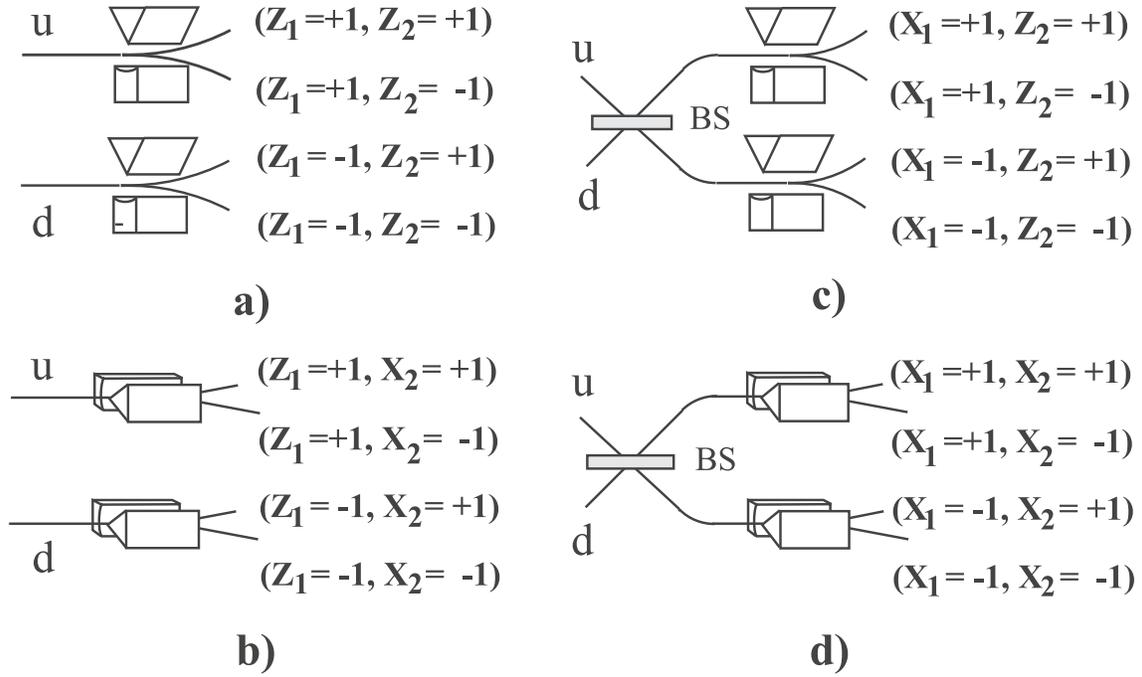}
\caption{  Devices for measuring pairs of the single-particle
observables of Eq. (\ref{obs}). A particle comes in from the left.
Note that in general the incoming states will have components in
both spatial modes $u$ and $d$ and of different spin. The devices
shown measure: a) $Z_1$ and $Z_2$; b) $Z_1$ and $X_2$; c) $X_1$
and $Z_2$; d) $X_1$ and $X_2$. BS in c) and d) stands for a
$50-50$ beam-splitter (see main text), which changes the basis of
path analysis from $|u\rangle, |d\rangle$, corresponding to a
measurement of $Z_1$, to $|u'\rangle, |d'\rangle$, thus leading to
a measurement of $X_1$. In a) and c) the Stern-Gerlach apparatus
are oriented along the $z$-axis (measurement of $Z_2$), in b) and
d) along the $x$-axis (measurement of $X_2$).}

\label{separate}
\end{figure}

To illustrate the physical meaning of the states and observables
in our scheme, we show in Fig. \ref{source} how a state such as
$|\psi_1\rangle$ in the form of Eq. (\ref{psi1}) can be prepared,
and in Fig. \ref{separate} we show the devices that measure pairs
of one-particle observables, such as $Z_1$ and $Z_2$.

\begin{figure}
    \centering
        \includegraphics[width= \columnwidth] {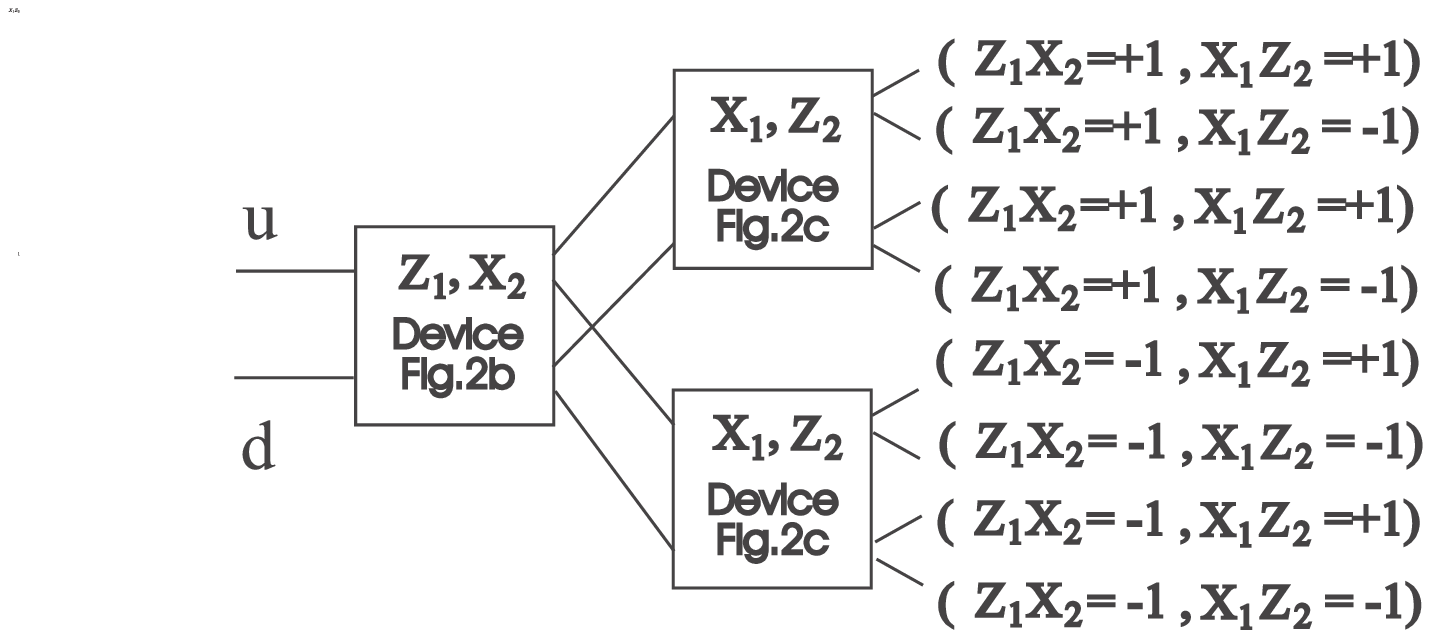}
        \caption{ Device for performing a joint measurement of $Z_1X_2$ and $X_1Z_2$.
         A device
        performing a joint measurement of $Z_1Z_2$ and $X_1X_2$ can be constructed in an analogous way.}
\label{joint}
\end{figure}

The devices of Figs. \ref{source} and \ref{separate} enable us to
realize step (i) of the protocol described above. As for step
(ii), Fig. \ref{joint} shows how a device performing a joint
measurement of $Z_1X_2$ and $X_1Z_2$ can be built out of the
building blocks of Fig. \ref{separate}.

Instead of leading to detectors, the outputs of the device of Fig.
\ref{separate}b, which measures $Z_1$ and $X_2$, are now connected
to two replicas of the device of Fig. \ref{separate}c, which
measure $X_1$ and $Z_2$. That the device indeed performs a joint
measurement of $Z_1X_2$ and $X_1Z_2$ can be demonstrated by
analyzing how it acts on the joint eigenstates of these two
observables.

Comparison with Fig. \ref{separate} shows that the first device
separates the two eigenspaces of the degenerate product observable
$Z_1X_2$. Eigenstates of $Z_1X_2$ with eigenvalue +1 are sent up,
those with eigenvalue -1 are sent down. It is important to note
that this is the only way in which {\it eigenstates} of $Z_1X_2$
are affected by the first device, i.e. they have exactly the same
form in terms of the two spatial modes leading to the respective
subsequent $X_1Z_2$-measuring device as they had in terms of the
modes entering the first device. One could say that the first
device ``almost'' performs an ideal Von Neumann measurement of the
observable $Z_1X_2$. The difference to a Von Neumann measurement
lies in the fact that the superposition between states with
$Z_1X_2=1$ and $Z_1X_2=-1$ is not destroyed by the device but only
made ineffective because the respective components of any incoming
state enter completely separated subsequent devices. Detection of
the particle behind one of those two subsequent devices is a Von
Neumann measurement of $X_1Z_2$ and at the same time completes the
measurement of $Z_1X_2$. As is evident from the structure of the
device of Fig. \ref{joint}, the measurement of $X_1Z_2$ is
performed by measuring $X_1$ and $Z_2$ separately as in Fig.
\ref{separate}c and multiplying the values.

While any device that performs a state analysis in the basis of
common eigenstates of $Z_1X_2$ and $X_1Z_2$ can be considered to
perform a joint measurement of these two observables, the
particular realization presented here has the merit of showing
explicitly that a joint measurement of two product observables is
performed, and how the information that could have been obtained
in the first stage of the measurement (the values of $Z_1$ and
$X_2$ separately) has to be partially erased in order to make the
second stage possible.

Let us now consider what happens when a particle in the state
$|\psi_1\rangle$ enters the device of Fig. \ref{joint}. Recall
from Eq. (\ref{sup}) that $|\psi_1\rangle$ is an equally-weighted
superposition of two states with opposite eigenvalues of $Z_1X_2$.
Therefore the particle has equal amplitudes for entering either of
the two $X_1Z_2$ devices. Explicit calculation confirms that the
particle can emerge only via one of those four outputs for which
the values of $Z_1X_2$ and $X_1Z_2$ are {\it opposite}. As
explained above, after it has been shown that $Z_1Z_2=X_1X_2=1$
for our particle source, NCHV predict exactly the complementary
set of outputs (for which $Z_1X_2$ and $X_1Z_2$ are equal).
Therefore the two theories give clearly conflicting predictions
for observable effects on a non-statistical level.
Of course, in a real experiment visibilities are
never perfect, and one would have to use some kind of inequality
to rigorously establish the contradiction.
(cf. \cite{ghz}).

The present scheme allows the simplest non-statistical
experimental test of non-contextuality that is known to us.
For a single-photon experiment that implements a statistical test
of NCHV versus QM see
\cite{michler}. Similarly to the original Kochen-Specker paradox
it requires only a single particle (though two degrees of
freedom). With the experimental setup consisting of a simple
interferometer, it shows particularly clearly that the appearance
of the paradox is related to the superposition principle.

\chapter{Hidden--Variable Theorems for Real Experiments}

\section{Introduction}

In the original derivation of hidden--variable theorems, such
as the Bell theorem, certain
idealizations were made. For example the detection efficiency was
originally assumed to be perfect. The case of non-unit efficiency
has since been treated in detail \cite{clauserhorne,clausershimony}.
As another idealization, the precision of the measurements
performed is usually not considered. When considering experimental
tests of the corresponding classes of hidden--variable theories, this appears to be an
important point for the following reason. An essential feature of
all the hidden--variable theorems is that observables
have to appear in different experimental contexts in order for a
contradiction to be obtained (i.e. observables have to be measured
simultaneously with different mutually exclusive observables).

For example, as we have seen above,
the Kochen--Specker theorem concerns trying to assign
values to all directions on the sphere subject to a constraint for
triads of orthogonal directions. One can only arrive at a
contradiction by considering several triads that have at least
some directions in common. For Kochen-Specker experiments this
implies that the observables corresponding to individual
directions (i.e. the squares of the spin components along these
directions) have to appear in different triads.

At first sight the usual derivations of hidden-variable theorems
seem to run into problems when the finite precision of real
experiments is taken into account, because then it seems
impossible to ascertain that the {\it same} observable is really
measured more than once in different experimental contexts. This
question seems to be of particular relevance for the
Kochen-Specker theorem in view of recent claims by Meyer that this theorem
is ''nullified'' when the measurements have only finite precision \cite{meyer}.

This claim was based on the fact that it is possible to assign values to all
rational directions of the sphere, which constitute a dense subset of all
directions. This construction was generalized by Kent \cite{kent}.
Meyer argued that, since by measurements with finite precision
one cannot discriminate a dense subset from its closure, this implies that
non-contextual hidden variables cannot be excluded by any real experiment of
the Kochen-Specker type. However, Meyer did not construct an explicit
non-contextual hidden-variable model for real experiments with finite
precision.

In the following we show how these questions can be resolved
by providing a general method for the derivation of
hidden-variable theorems for real experiments. In order to achieve
this the concept of observable has to be changed in such a way
that it has an {\it operational} meaning. For concreteness,
imagine that an observer wants to perform a measurement of the
spin square along a certain direction $\vec{n}$. There will be a
certain experimental procedure for trying to do this as accurately
as possible. We will refer to this procedure by saying that he
sets the ''control switch'' of his apparatus to the position
$\vec{n}$. In all experiments that we will discuss only a finite
number of different switch positions is required. By definition
different switch positions are clearly distinguishable for the
observer, and the switch position is all he knows about.
Therefore, in an operational sense the measured physical
observable is entirely defined by the switch position. From the
above definition it is clear that the same switch position can be
chosen again and again in the course of an experiment.

In general one has to allow for the possibility that the switch
position $\vec{n}$ does not uniquely determine the physical state
of the measuring apparatus, i.e. there may be (hidden) properties
of the apparatus over which the observer does not have full
control but which may influence the result of any given
measurement. Following the philosophy of deterministic hidden
variable theories, one therefore has to assume
that the result of any measurement will be determined not only by
the hidden properties of the system, but also by those of the
measuring apparatus.

In the present paper we do not discuss stochastic hidden variable
theories explicitly. This does not limit the generality of the results derived
because the existence of a stochastic hidden variable model for a given
physical system implies that also an underlying deterministic model can be
constructed which reproduces the probabilities of the stochastic model.
Therefore e.g. ruling out all possible non-contextual deterministic
hidden-variable models implies ruling out all possible non-contextual
stochastic models as well.

\section{Kochen--Specker Theorem for Real Experiments}

As a concrete application of the ideas expressed in the two
preceding paragraphs, we are now going to show how non-contextual
hidden variables can be excluded by real experiments. Let us note
that local hidden variables can be ruled out using an equivalent
approach.

In the original Kochen-Specker situation one considers a spin-1
particle. In the ideal case of perfect precision, the relevant
observables are the squares of the spin components, denoted by
$S^2_{\vec{n}}$ for arbitrary directions $\vec{n}$. For a spin-1
particle one has
\begin{equation}
S^2_{\vec{n}_1}+S^2_{\vec{n}_2}+S^2_{\vec{n}_3}=2 \label{sum}
\end{equation}
for every triad of orthogonal directions
$\{\vec{n}_1,\vec{n}_2,\vec{n}_3\}$. As the possible results for
every $S^2_{\vec{n}_i}$ are 0 or 1, this implies that in the ideal
case for every measurement of three orthogonal spin squares two of
the results will be equal to one, and one of them will be equal to
zero.

Let us emphasize that in our approach the operational observables
are defined by the switch positions (i.e. by the best effort and
knowledge of the experimenter) and therefore are not exactly
identical to the exact quantum mechanical observables. In the
following the symbol $S^2_{\vec{n}}$ will denote the operational
observable defined by the switch position $\vec{n}$, and the term
direction will be used as a synonym for switch position.

In a deterministic hidden variable theory one assumes that for
every individual particle the result of the measurement of any
observable $S^2_{\vec{n}}$ is predetermined by hidden properties.
In {\it non-contextual} hidden variable theories it is furthermore
assumed that this predetermined result does not depend on the
''context'' of the measurement, in particular which other
observables are measured simultaneously with $S^2_{\vec{n}}$, but
only on the switch position $\vec{n}$ and the hidden variables.

In the ideal case one could define non-contextuality in such a way that the
predetermined value of some quantum mechanical observable $X$ is required to be
independent of the simultaneously measured observables only if they exactly
commute with $X$. Note that only in the ideal case the observables
corresponding to precise directions would have an operational meaning. It is
evident that this weaker form of non-contextuality can only be tested in the
idealized case of infinite precision.

In general the result may depend both on the hidden properties of
the system and of the apparatus. Let us denote the hidden
variables of the system by $\lambda$ and those of the apparatus by
$\mu$. For further use, let us denote the ensemble of all possible
pairs $(\lambda,\mu)$ by $\Lambda$. As explained above, the
philosophy of non-contextual hidden variables implies the
existence of a function $S^2_{\vec{n}}(\lambda,\mu)$ taking values
0 and 1 which describes the result of a measurement with switch
position $\vec{n}$ on a system characterized by $\lambda$ with an
apparatus characterized by $\mu$. For fixed $\lambda$ and $\mu$
this function therefore assigns a value 0 or 1 to the switch
position $\vec{n}$.
Let us note that the models discussed by Clifton and Kent \cite{clifton}
are not non-contextual in the present sense because in these
models the result of a measurement of $S^2_{\vec{n}}$ in general does not only
depend on $\lambda$, $\mu$, and $\vec{n}$, but also on the other observables
measured simultaneously.

A Kochen--Specker experiment can now be performed by testing the
validity of Eq. (\ref{sum}) for a judiciously chosen set of triads
of directions. Therefore the apparatus is required to have three
switches where the three directions of a given triad can be
chosen. Because the switch positions do not correspond to the
ideal quantum mechanical observables the sum of the three results
will not always be equal to 2. Nevertheless a contradiction
between non-contextuality and quantum mechanics can be obtained in
the following way.

From the Kochen-Specker theorem it follows that there are finite
sets of triads for which no value assignment consistent with Eq.
(\ref{sum}) is possible \cite{ks,peresks}. Let us choose such a
Kochen-Specker set of triads
\begin{equation}
\left\{ \{\vec{n}_1,\vec{n}_2,\vec{n}_3\},
\{\vec{n}_1,\vec{n}_4,\vec{n}_5\}, ..., \right\}. \label{ksset}
\end{equation}
Let us emphasize that at least some of the switch positions
$\vec{n}_i$ have to appear in several of the triads (clearly
otherwise there could be no inconsistency). Let us denote the
number of triads in the Kochen-Specker set (\ref{ksset}) by $N$.
The set is constructed in such a way that if one can show for some
fixed $\lambda$ and $\mu$ that
\begin{equation}
S^2_{\vec{n}_i}(\lambda,\mu)+S^2_{\vec{n}_j}(\lambda,\mu)+S^2_{\vec{n}_k}(\lambda,\mu)=2
\label{sumlambda}
\end{equation}
is valid for $N-1$ of the triads
$\{\vec{n}_i,\vec{n}_j,\vec{n}_k\}$, one obtains the prediction
that it has to be violated for the final triad.

Suppose that for the first triad
$\{\vec{n}_1,\vec{n}_2,\vec{n}_3\}$ in the Kochen-Specker set one
finds that the sum of the results is equal to 2 in a fraction
greater than $1-\epsilon$ of all cases. For the hidden variables
this implies that
\begin{equation}
S^2_{\vec{n}_1}(\lambda,\mu)+S^2_{\vec{n}_2}(\lambda,\mu)+S^2_{\vec{n}_3}(\lambda,\mu)=2
\end{equation}
for all $(\lambda,\mu) \in \Lambda_1$, where $\Lambda_1$ is some
subset of the set of all hidden variables $\Lambda$ with measure
$p(\Lambda_1)\geq 1-\epsilon$ (by definition $p(\Lambda)=1$).
Suppose furthermore that one establishes in the same way for the
second triad $\{\vec{n}_1,\vec{n}_4,\vec{n}_5\}$ that
\begin{equation}
S^2_{\vec{n}_1}(\lambda,\mu)+S^2_{\vec{n}_4}(\lambda,\mu)+S^2_{\vec{n}_5}(\lambda,\mu)=2
\end{equation}
for all  $(\lambda,\mu) \in \Lambda_2$ with $p(\Lambda_2)\geq
1-\epsilon$ where in general $\Lambda_2$ is a different subset of
$\Lambda$, and so on for all $N-1$ triads except the final one.

This implies that for all $(\lambda,\mu)$ in the intersection of
sets $\Lambda_{\cap} :=  \Lambda_1 \cap \Lambda_2 \cap ... \cap
\Lambda_{N-1}$ the sum of results is equal to 2. Consequently,
because of the structure of the Kochen-Specker set the sum of the
results for the final triad has to be different from 2 (i.e. 0, 1
or 3) for all pairs $(\lambda,\mu) \in \Lambda_{\cap}$. This leads
to the experimental prediction that the sum of results will be
different from 2 for the final triad in a fraction
$p(\Lambda_{\cap})$ of all cases. From the property of
sub-additivity $(p(\cup_i A_i) \leq \sum_i p(A_i))$ of the measure
$p$ it immediately follows that
\begin{equation}
p(\Lambda_{\cap}) \geq 1 - (N-1) \epsilon. \label{ineq}
\end{equation}

Therefore in order to experimentally disprove non-contextual
hidden variables one only needs to show that the sum of results is
equal to 2 in a fraction of all cases that is greater than $(N-1)
\epsilon$.

If we assume for simplicity that $\epsilon$ is defined such that
the fraction of ''correct'' (equal to 2) results is larger than
$1-\epsilon$ for all triads (including the final one) then the
above results allow us to derive a bound on the size of the
experimental imperfection $\epsilon$ such that an experimental
contradiction with non-contextuality can still be obtained:
$\epsilon$ has to be smaller than $1/N$. Note that $\epsilon$
describes all the imperfections of a real experiment including
finite precision but also e.g. imperfect state preparation and
non-unit detection efficiency. The value of $N$ and therefore of
the bound on $\epsilon$ depends on the particular Kochen-Specker
set used \cite{ks,peresks}.

As we have already noted above, an inevitable requirement for the
contradiction to be obtained is the fact that the function
$S^2_{\vec{n_1}}(\lambda,\mu)$, or in general functions
corresponding to at least some switch positions, appear in more
than one out of the $N$ triads. This appearance of the same
function in different lines of the mathematical proof
(corresponding to different experimental contexts) is possible in
spite of finite experimental precision only because we defined our
observables operationally via the switch positions.

We have shown how non-contextual hidden-variable theories can be
disproved by real experiments. This clarifies
questions raised by \cite{meyer}. In view of our results, we would
assert that the Kochen-Specker theorem is not ''nullified'' by
finite measurement precision. Let us note that independent
arguments in favor of this conclusion were given in
\cite{merminfp,apple,apple2,larsson}. Our suggestion how to perform a
Kochen-Specker experiment was inspired by some of Mermin's remarks
in \cite{merminfp}.

Using the same method one can also show that local
hidden variables can be disproved in real experiments, e.g. using
the GHZ \cite{ghz} form of Bell's theorem which is also based on
sets of propositions that cannot be consistently satisfied by
hidden variables. Inequalities analogous to Eq. (\ref{ineq}) can
be derived and tested experimentally \cite{pan}.

\section{Hidden Variables: Perspectives}

In the previous chapter we have presented a very simple Kochen--Specker type argument.
It seems unlikely that much further simplification is possible.
In this chapter we have analyzed the derivation of hidden--variable theorems
for real experimental conditions. We have seen that the theorems,
including those on non--contextuality, are robust under real--world
conditions and thus experimentally testable.

Let us emphasize that 36 years after Bell (and 68 years after von Neumann)
there are still interesting open questions in the field of hidden variables. Most importantly, on
the experimental side, a loophole--free
demonstration of the violation of Bell's inequalities is still missing.
Such an experiment would require both space-like separation of the measurements
performed on each entangled pair and high detection efficiency.
For the realization of such experiments new ways of establishing
contradictions between local hidden variables and quantum mechanics
may be helpful. E.g. Eberhard \cite{eberhard} showed that the required
detection efficiencies are lower if non--maximally entangled states
together with appropriate analyzer settings are used.
A possible new approach would be to consider adaptive measurements
\cite{rudolph}, or in general joint measurements on several pairs.

On the theoretical side, there are many connections between the study
of general Bell's inequalities and the classification and quantification
of entanglement, cf. e.g. \cite{scarani}. A particularly interesting
open question is whether the so--called ``bound entangled'' states
\cite{boundent}, which are states from which no maximal entanglement can
be distilled, admit local hidden variable models, i.e. whether
there is entanglement without non--locality.

One may also hope that a detailed understanding of the
quantum weirdness, i.e. the differences between quantum mechanics
and the classical world view, might help to generate ideas
how to exploit it in order to perform tasks that are classically
impossible.

\addtocontents{toc}{\contentsline {chapter}{}{}}

\addchap{Conclusions and Outlook}

During the three years of my PhD studies I had the good fortune to
come into contact with many fields, some of which are not even
mentioned in this thesis. Quite naturally I learned about various
sub-fields of quantum information, ranging from cloning and
quantum cryptography, over quantum computing, to the study of
entanglement and its purification. I got to know quantum optics,
another field which had not figured in my undergraduate studies,
from the experimental and also somewhat from the theoretical side.
I learned a lot about hidden-variable theorems and the related
experiments and some basic but important facts about the practical
aspects of decoherence, mostly from my experimental colleagues.
Towards the end I was glad to learn some things about entanglement
in quantum field theory, a topic I liked because it created a
bridge to my earlier studies. I also found that my expectations
had been correct: indeed there were many opportunities to
discuss, learn and think about the basic questions of quantum
physics, together with Anton Zeilinger, \v{C}aslav Brukner, and
many others.

One of the most important things that I learned is that it is very
nice and that it can also be quite fruitful for a theorist to be
in close (in my case: permanent) contact with an experimental
group, especially if it is such a good one. Not least because your
knowledge of physics is constantly tested by the questions of your
experimental friends, which have a tendency to always be related
to the real world. I learned other things which I think will be useful, such as writing a proposal, organizing a
workshop. Once, we even made a movie. All this was usually done in
a team of great people, from about ten different nations. So much
for myself.

What have we learned from quantum information in general? Most
importantly, that, with the help of quantum physical systems, one
can do things that are unthinkable classically, starting of course
with the discovery of Bell's inequalities. In the last years we
have also learned a lot about what can be done in practice in the
lab. This includes many amazing things, such as multi-particle
entanglement \cite{ghzexp}, the interference of large molecules
\cite{c60}, the study of single ions in traps coupled via single
phonons \cite{monroe}, and single photons interacting with
single atoms in cavities of incredibly high quality
\cite{flythru}. It is probably also fair to say that quantum
information has led to a new way of looking at physics, for
example we see entanglement almost everywhere.

Let me try to summarize the major challenges for quantum
information at the present stage by two questions: What else could
we do (with quantum systems), and what can we really build? The
first question has a theoretical, the second an experimental
flavor, but people from both sides are trying to find answers to
both. A natural way of attacking the first one is to look for new
quantum algorithms that out-perform classical ones. Physicists and
computer scientists are also investing a lot of effort into trying
to prove general results on the power of quantum computation. One
may also feel that "quantum non-locality" has not yet been fully
exploited. There must be more that one can do with distributed
entanglement than violate Bell's inequalities and perform
cryptography and teleportation. One promising result is the
reduction of communication complexity \cite{commcomplexity}.

As for the second question, the main goals are to build a quantum
computer of serious size and to achieve quantum communication over
long distances. Hope rests on the continuous improvement of solid
state techniques \cite{kane}, for example for semi-conductors and
super-conductivity, and also in new technological achievements,
such as Bose-Einstein condensation \cite{bec} and the development of
laser cooling \cite{cooling}. The main difficulty
in designing and building a quantum computer is to keep quantum
coherence, where normally it is lost very fast. This is attempted
using a combination of technological approaches such as cooling,
isolation, the use of systems which have low decoherence by
nature, and algorithmic methods, such as quantum error correction
and fault-tolerant computing.

Thus, there is a close connection between quantum computing and a
more foundationally oriented research programme, the preparation
and study of larger and larger superposition states,
``Schr\"{o}dinger cats''. One of the driving hopes behind such a
programme was formulated in the preface to this thesis: something
new could turn up. We might find the limits of validity of quantum
physics, maybe even something like the border of the classical
world. At present such hopes, although no strangers to the
author's heart, seem preposterous if not unreasonable. One
argument in their favor which has some appeal for the author is
that usually in physics linearity is an approximation. The
connection between the linearity of quantum mechanics and special
relativity should be kept in mind in this context, but it is well
known that non-linear modifications of quantum physics are
conceivable if some of its basic assumptions are given up.

There are other more modest hopes for the future of quantum
information. There is little doubt that we will learn a lot more
about physics in a practical and quantitative sense. We also hope
for new conceptual insights, for example concerning the relation
of quantum and classical information. Quantum information should
meet other areas of physics besides quantum optics, most
prominently statistical physics and quantum field theory. New
interesting physics should come out of such encounters. A more
ambitious hope again is that the concept of information may help
us to arrive at a deeper understanding of the basic principles of
quantum physics \cite{bruknerprl,bruknerdiss,antonfp}.

Personally I hope to use many of the things that I have learned
and to continue working in a fascinating field dealing with
fundamental questions and yet close to the real world of
experiments.

\backmatter

\chapter{Papers by the Author}

\begin{enumerate}
\addcontentsline{toc}{section}{List of Publications}
\item
G. Weihs, T. Jennewein, C. Simon, H. Weinfurter, and A. Zeilinger:
{\em Violation of Bell's inequality under strict Einstein locality
conditions.} Phys. Rev. Lett., {\bf 81}, 5039-43 (1998), quant-ph/9810080

\item
C. Simon, G. Weihs, and A. Zeilinger: {\em Quantum Cloning and
Signaling.} Acta Phys. Slov., {\bf 49}, 755-760 (1999)

\item
C. Simon, G. Weihs, and A. Zeilinger: {\em Optimal quantum cloning and
universal NOT without quantum gates,} J. Mod. Opt., {\bf 47}, 233-246
(2000)

\item
C. Simon, G. Weihs, and A. Zeilinger: {\em Optimal Quantum Cloning via
Stimulated Emission,} Phys. Rev. Lett., {\bf 84}, 2993 (2000),
quant-ph/9910048

\item
T. Jennewein, C. Simon, G. Weihs, H. Weinfurter, and A. Zeilinger:
{\em Quantum Cryptography with Polarization Entangled Photons,} Phys.
Rev. Lett., {\bf 84}, 4729 (2000), quant-ph/9912117

\item
J. Kempe, C. Simon, and G. Weihs: {\em Optimal Photon Cloning,} Phys.
Rev. A {\bf 62}, 032302 (2000), quant-ph/0003025

\item
C. Simon, M. \.{Z}ukowski, H. Weinfurter, and A. Zeilinger: {\em A feasible
"Kochen-Specker" experiment with single particles,} Phys. Rev.
Lett., {\bf 85}, 1783 (2000), quant-ph/0009074
\end{enumerate}

\chapter{Acknowledgements}

There are many people to whom I feel grateful and who I would like
to thank at this occasion.

Meinen Eltern Hannelore und G\"{u}nter Simon f\"{u}r die
langj\"{a}hrige finanzielle und moralische Unterst\"{u}tzung und
daf\"{u}r, daß sie nie den geringsten Druck auf mich ausge\"{u}bt
haben. Außerdem meiner Mutter daf\"{u}r, daß sie mir beigebracht
hat, wie man Dinge herausfinden kann, die man nicht weiß, und
meinem Vater f\"{u}r seinen Humor und seinen immer beruhigenden
Einfluß.

Anton Zeilinger, my ``doctor-father'', as we say in German, for
his moral and financial support through the years of my PhD, for
providing me with an enormous range of opportunities, for
accepting my change from experiment to theory with the utmost
tolerance, and for his understanding and sympathy for my problems
with my hands. Anton's deep insight and love for physics, his
honesty and his striving for highest quality in all undertakings
will remain exemplary for me wherever I will go.

All my collaborators, colleagues, teachers and friends for many
enjoyable and instructive discussions about physics and life. I am
grateful also to those who are not mentioned by name in the
following. In particular let me thank:

Gregor Weihs for introducing me to experimental physics (together
with Thomas) and for our common cloning adventure.

\v{C}aslav Brukner for our daily life together since we share an
office here in Vienna, for his sensitivity and help in difficult
phases, and for bearing with me all this time at an average
distance of 30 centimeters.

Marek \.{Z}ukowski for all the books he gave me.

Vladimir Bu\v{z}ek for his unfailing positive attitude which
remounted my morale more than once, and for reading this thesis.

Jian-Wei Pan for our common work, which is not part of this
thesis, and for teaching me a lot about China.

Dik Bouwmeester for inviting me to join him in Oxford.

Thomas Jennewein for many open-air lunches together in Innsbruck.

Olaf Nairz for many packets of ``Manner Schnitten'', and Alois
Mair for a lot of chocolate in all forms.

Gerbrand van der Zouw for the cactus.

Julia Petschinka and Guido Czeija for some enlightenment.

Helmut Neufeld, Walter Grimus and Gerhard Ecker for guiding my
very first steps into theoretical physics.

Peter Stuparits, my physics teacher in high school, for conveying
his enthusiasm about quantum physics.

Reinhold Bertlmann for his seminar together with Anton, and for
consenting to be my opponent.

Larissa Cox for her big help in typing all this and for
straightening out my continental English.

Jakob Kellner for his friendship through many years, for having
always forced me to think clearly about physics by posing many
excellent questions, usually while feeding me, and for his
enormous help in preparing this thesis.

Julia Kempe for her long-lasting friendship, her good example in
many things, and for asking me about Lambda atoms.

Beate Stengg for her patience and help during my last years of
school and first years of study, which were sometimes difficult,
as she knows.

\chapter{Curriculum Vitae}

{\Large Christoph Simon}

\begin{tabular}{l p{13cm}}
1974       & Born in Oberwart (Burgenland, Austria) on July 25.\\
1980-1984  & Elementary school (Volksschule) in Pinkafeld.\\
1984-1992  & High school (realistisches Gymnasium) in Obersch\"{u}tzen\\
1992-1996  & Studied physics at the University of Vienna.
Specialization in theoretical
                elementary particle physics.\\
1995       & Summer student at CERN, Geneva.\\
1996-1997  & Obtained the Dipl\^{o}me d'Etudes Approfondies (DEA)
de Physique Théorique at the Ecole Normale Supérieure in Paris.\\
1997-2000  & PhD student in the group of Prof. Anton Zeilinger at the University of
Innsbruck and the University of Vienna.
Participation in experiments on Bell's inequality and quantum
cryptography. Theoretical work on quantum cloning by stimulated emission,
the no-signaling
condition and quantum dynamics, hidden-variable theorems, and
entanglement purification.\\
\end{tabular}


\begin{thebibliography}{99}
\bibitem{bec} M.H.      Anderson, J.R. Ensher, M.R. Matthews, C.E. Wieman, and E.A. Cornell, Science {\bf 269}, 198 (1995).
\bibitem{apple2} D.M.   Appleby, quant-ph/0005010.
\bibitem{apple} D.M.    Appleby, quant-ph/0005056.
\bibitem{c60} M.        Arndt, O. Nairz, J. Voss-Andreae, C. Keller, G. van der Zouw, and A. Zeilinger, Nature {\bf 401}, 680 (1999).
\bibitem{aspect}A.      Aspect, J. Dalibard, and G. Roger, Phys. Rev. Lett. {\bf 49}, 1804 (1982).
\bibitem{bechmann}H.    Bechmann-Pasquinucci and N. Gisin, Phys. Rev. A {\bf 59}, 4238 (1999).
\bibitem{bell} J.S.     Bell, Physics (Long Island City, N.Y.) 1, 195 (1964). Reprinted in \cite{bellbook}.
\bibitem{bellks} J.S.   Bell, Rev. Mod. Phys. {\bf 38}, 447 (1966).
\bibitem{bellbook} J.S. Bell, {\it Speakable and Unspeakable in Quantum Mechanics} (Cambridge Univ. Press, Cambridge, 1987).
\bibitem{teleport}C.H.  Bennett, G. Brassard, C. Cr\'{e}peau, R. Jozsa, A. Peres, and W. K. Wootters, Phys. Rev. Lett. {\bf 70}, 1895 (1993).
\bibitem{crypto}C.H.    Bennett, G. Brassard, S. Briedbart, and S. Wiesner, Advances in Cryptology: Proceedings of Crypto '82 (Plenum, New York), 267 (1983).
\bibitem{qi}C.H.        Bennett and D.P. DiVincenzo, Nature {\bf 404}, 247 (2000).
\bibitem{telep} D.      Bouwmeester, J.W. Pan, K. Mattle, M. Eibl, H. Weinfurter, and A. Zeilinger, Nature {\bf 390}, 575 (1997).
\bibitem{Dik}D. Bouwmeester, private communication.
\bibitem{braunstein} S.L.       Braunstein, V. Bu\v{z}ek, and M. Hillery, quant-ph/0009076.
\bibitem{bruknerprl} C. Brukner and A. Zeilinger, Phys. Rev. Lett. {\bf 83}, 3354 (1999).
\bibitem{bruknerdiss} C.        Brukner, {\it Information in individual quantum systems} (PhD thesis, Technical University of Vienna, 1999).
\bibitem{brussprl} D.   Bru\ss, A. Ekert, and C. Macchiavello, Phys. Rev. Lett. {\bf 81}, 2598 (1998).
\bibitem{buzeknot} V.   Bu\v{zek}, M.~Hillery, and R.F.Werner, {\em Phys. Rev. A} {\bf 60}, R2626 (1999); {\em J. Mod. Opt.} {\bf 47}, 211 (2000).
\bibitem{buzbraun} V.   Bu\v{z}ek, S.L. Braunstein, M. Hillery, and D. Bru\ss, Phys. Rev. A {\bf 56}, 3446 (1997).
\bibitem{buzekhillery}  V.      Bu\v{z}ek and M. Hillery, Phys. Rev. A {\bf 54}, 1844 (1996).
\bibitem{buzekprl}  V.  Bu\v{z}ek and M. Hillery, Phys. Rev. Lett. {\bf 81}, 5003 (1998).
\bibitem{ddimgen}       C. Simon and V. Bu\v{z}ek, in preparation
\bibitem{cab} A.        Cabello and G. Garc\'{\i}a-Alcaine, Phys. Rev. Lett. {\bf 80}, 1797 (1998).
\bibitem{cald}A.R.      Calderbank and P. W. Shor, Phys. Rev. A {\bf 54}, 1098 (1996).
\bibitem{cirac}J.I.     Cirac and P. Zoller, Phys. Rev. Lett. {\bf 74}, 4091 (1995).
\bibitem{clausershimony} J.F.   Clauser and A. Shimony, Rep. Prog. Phys. {\bf 41}, 1881 (1978).
\bibitem{clauserhorne} J.F.     Clauser and M. A. Horne, Phys. Rev. D {\bf 10}, 526 (1974).
\bibitem{commcomplexity} R.     Cleve and H. Buhrmann, Phys. Rev. A {\bf 56}, 1201 (1997).
\bibitem{clifton} R.    Clifton and A. Kent, P. Roy. Soc. Lond. A MAT {\bf 456} 2101 (2001)
\bibitem{Collett} M.J.  Collett, Phys. Rev. A {\bf 38}, 2233 (1988).
\bibitem{nmr}D.G.       Cory et al, quant-ph/0004104.
\bibitem{czachor} M.    Czachor, Phys. Rev. A {\bf 49}, 2231 (1994).
\bibitem{Czachor97}M.   Czachor and M. Kuna, Phys. Rev. A {\bf 58}, 128 (1998).
\bibitem{DeM} F.        De Martini, Phys. Rev. Lett. {\bf 81}, 2842 (1998).
\bibitem{mussi} F.      De Martini and V. Mussi, Fortschr. Phys. {\bf 48} (5-7) 413 (2000)
\bibitem{dieks} D.      Dieks, Phys. Lett. {\bf 92}A, 271 (1982).
\bibitem{eberhard}P. H. Eberhard, Phys. Rev. A {\bf 47}, R747 (1993)
\bibitem{singlephot} S. Friberg, C.K. Hong, and L. Mandel, Phys. Rev. Lett. {\bf 54}, 2011 (1985).
\bibitem{ghirardi} G.C. Ghirardi, A. Rimini, and T. Weber, Lett. Nuov. Cim. {\bf 27}, 293 (1980).
\bibitem{gisinhpa} N.   Gisin, Helv. Phys. Acta {\bf 62}, 363 (1989).
\bibitem{gisinweinberg} N.      Gisin, Phys. Lett. A {\bf 143}, 1 (1990).
\bibitem{gisinbound}  N.        Gisin, Phys. Lett. A {\bf 242}, 1 (1998).
\bibitem{gisinmassar} N.        Gisin and S. Massar, Phys. Rev. Lett. {\bf 79}, 2153 (1997).
\bibitem{gleason}A.M.   Gleason, J. Math. Mech. {\bf 6}, 885 (1957).
\bibitem{ghzajp} D.M.   Greenberger, M. Horne, A. Shimony, and A. Zeilinger, Am. J. Phys. {\bf 58}, 1131 (1990).
\bibitem{ghz} D.M.      Greenberger, M. Horne, and A. Zeilinger, in {\it Bell's Theorem, Quantum Theory, and Conceptions of the Universe}, edited by M. Kafatos (Kluwer, Dordrecht, 1989).
\bibitem{grover}L.K.    Grover, Phys. Rev. Lett. {\bf 79}, 325 (1997).
\bibitem{flythru} E.    Hagley, X. Maitre, G. Nogues, C. Wunderlich, M. Brune, J.M. Raimond, and S. Haroche, Phys. Rev. Lett. {\bf 79}, 1 (1997).
\bibitem{herbert} N.    Herbert, Found. Phys. {\bf 12}, 1171 (1982).
\bibitem{boundent}M.    Horodecki, P. Horodecki, and R. Horodecki, Phys. Rev. Lett. {\bf 80}, 5239 (1998).
\bibitem{hughston}L.P.  Hughston, R. Jozsa, W. K. Wootters, Phys. Lett. A {\bf 183}, 14 (1993).
\bibitem{qft}C. Itzykson and J.--B. Zuber, {\em Quantum Field Theory} (McGraw--Hill, 1985).
\bibitem{jennewein}T.   Jennewein, C. Simon, G. Weihs, H. Weinfurter, A. Zeilinger, Phys. Rev. Lett {\bf 84}, 4729 (2000).
\bibitem{kane} B.E.     Kane, Nature {\bf 393}, 133 (1998).
\bibitem{kent} A.       Kent, Phys. Rev. Lett. {\bf 83} (1999) 3755.
\bibitem{keylwerner} M. Keyl and R.F. Werner, J. Math. Phys. {\bf 40}, 3283 (1999).
\bibitem{ks} S. Kochen and E. P. Specker, J. Math. and Mech. {\bf 17}, 59 (1967).
\bibitem{Source} P.G.   Kwiat, K. Mattle, H. Weinfurter, A. Zeilinger, A.V. Sergienko, and Y. Shih, Phys. Rev. Lett. {\bf 75}, 4337 (1995).
\bibitem{larsson} J.--\AA.      Larsson, quant-ph/0006134.
\bibitem{mandel} L.     Mandel, Nature {\bf 304}, 188 (1983).
\bibitem{massarpop}S.   Massar and S. Popescu, Phys. Rev. Lett. {\bf 74}, 1259 (1995).
\bibitem{merminfp} N.D. Mermin, quant-ph/9912081.
\bibitem{cooling} H.J.  Metcalf and P. van der Straten, {\it Laser Cooling and Trapping} (Springer-Verlag, New York, 1999).
\bibitem{meyer} D.      Meyer, Phys. Rev. Lett. {\bf 83}, 3751 (1999).
\bibitem{michler} M.    Michler, H. Weinfurter, and M. \.{Z}ukowski, Phys. Rev. Lett. {\bf 84}, 5457 (2000).
\bibitem{Migdall} A.    Migdall, Phys. Today, January 1999, p. 41 (1999).
\bibitem{milonni} P.W.  Milonni and M.L. Hardies, Phys. Lett. {\bf 92}A, 321 (1982).
\bibitem{monroe}C.      Monroe, D.M. Meekhof, B.E. King, W.M. Itano, and D.J. Wineland, Phys. Rev. Lett. {\bf 75}, 4714 (1995).
\bibitem{cerf}  N.J. Cerf, Acta Phys. Slov. {\bf 48}, 115 (1998).
\bibitem{naik}D.S.      Naik, C.G. Peterson, A.G. White, A.J. Berglund, P.G. Kwiat, Phys. Rev. Lett {\bf 84}, 4733 (2000).
\bibitem{ghzexp} J.W.   Pan, D. Bouwmeester, M. Daniell, H. Weinfurter, and A. Zeilinger, Nature {\bf 403}, 515 (2000).
\bibitem{pan} J.W.      Pan, D. Bouwmeester, M. Daniell, H. Weinfurter, and A. Zeilinger, Nature {\bf 403}, 515 (2000).
\bibitem{peresks} A.    Peres, J. Phys. A {\bf 24}, L175 (1991).
\bibitem{peresbook} A.  Peres, {\it Quantum Theory: Concepts and Methods} (Kluwer Academic Publishers, Dordrecht, The Netherlands, 1993).
\bibitem{qmfromns}S.    Popescu and D. Rohrlich, Found. Phys. {\bf 24}, 379 (1994).
\bibitem{preskill}J.    Preskill, Lecture Notes on Quantum Computation, \verb!http://www.theory.caltech.edu/people/preskill/ph229/#lecture!
\bibitem{rudolph}T.     Rudolph, private communication.
\bibitem{schwinger} J.J.        Sakurai, {\it Modern quantum mechanics} (Addison-Wesley, 1994), Section 3.8.
\bibitem{scarani}V.     Scarani and N. Gisin, private communication.
\bibitem{shor2}P.W.     Shor, Phys. Rev. A {\bf 52}, 2493 (1995).
\bibitem{faultt}P.W.    Shor, Proc. 37th Symp. on Foundations of Computer Science.
\bibitem{shor}P.W.      Shor, SIAM J. Comp. {\bf 26}, No. 5, 1484 (1997).
\bibitem{simbuzgis}C.   Simon, V. Bu\v{z}ek, and N. Gisin, in preparation.
\bibitem{specker} E.P.  Specker, {\it Selecta} (Birkh\"{a}user Verlag, Basel, 1990).
\bibitem{steane}A.M.    Steane, Phys. Rev. Lett. {\bf 77}, 793 (1996).
\bibitem{tittel}W.      Tittel, J. Brendel, H. Zbinden, M. Gisin, Phys. Rev. Lett. {\bf 84}, 4737 (2000).
\bibitem{cqed}Q.A.      Turchette, C.J. Hood, W. Lange, H. Mabuchi, and H.J. Kimble, Phys. Rev. Lett. {\bf 75}, 4710 (1995).
\bibitem{Walls} D.F.    Walls and G.I. Milburn, {\it Quantum Optics} (Springer-Verlag, Berlin, 1995), Chap. 5.
\bibitem{weihs}G.       Weihs, T. Jennewein, C. Simon, H. Weinfurter, and A. Zeilinger, Phys. Rev. Lett. {\bf 81}, 5039 (1998).
\bibitem{werner}  R.F.  Werner, Phys. Rev. A {\bf 58}, 1827 (1998).
\bibitem{wzurek} W.K.   Wootters and W.H. Zurek, Nature (London) {\bf 299}, 802 (1982).
\bibitem{kimble} J.     Ye, D.W. Vernooy, and H.J. Kimble, Phys. Rev. Lett. {\bf 83}, 4987 (1999)
\bibitem{antonfp} A.    Zeilinger, Found. Phys. {\bf 29}, 631 (1999).
\bibitem{ghzvorschlag} A.       Zeilinger, M.A. Horne, H. Weinfurter, and M. \.{Z}ukowski, Phys. Rev. Lett. {\bf 78}, 3031 (1997).
\bibitem{azent}A.       Zeilinger and M. \.{Z}ukowski, to appear in Rev. Mod. Phys.
\bibitem{marek} M.      \.{Z}ukowski, Phys. Lett. A {\bf 157}, 198 (1991).
\bibitem{decoherence}W.H.       Zurek, Physics Today {\bf 44}(10), October, 36 (1991).
\end{thebibliography}
\end{document}